\documentclass[twocolumn,showpacs,pre,aps,longbibliography,superscriptaddress,nofootinbib]{revtex4-2}
\usepackage[bookmarks=false,linkcolor=blue,urlcolor=blue,colorlinks,citecolor=blue]{hyperref}
\pdfoutput=1
\usepackage[english]{babel}

\usepackage{quantikz}
\usepackage{adjustbox}
\usepackage{pifont}
\usepackage{multirow}

\usepackage{amsmath}
\usepackage{amssymb, amsthm, bm, bbm}
\usepackage{mathtools}

\usepackage{graphicx}
\usepackage{subfigure}
\usepackage{relsize}
\graphicspath{{figures/}}

\usepackage{color}
\definecolor{pink}{rgb}{1,0,0.9}

\newtheorem{theorem}{Theorem}

\newtheorem{corollary}{Corollary}

\begin{document}

\title{Entanglement scaling in matrix product state representation of smooth functions and their shallow quantum circuit approximations}

\author{Vladyslav Bohun$^*$}
\affiliation{Haiqu, Inc., 95 Third Street, San Francisco, CA 94103, USA}
\author{Illia Lukin$^*$}
\affiliation{Haiqu, Inc., 95 Third Street, San Francisco, CA 94103, USA}
\affiliation{Akhiezer Institute for Theoretical Physics, NSC KIPT, Akademichna 1, 61108 Kharkiv, Ukraine}
\author{Mykola Luhanko}
\affiliation{Haiqu, Inc., 95 Third Street, San Francisco, CA 94103, USA}
\affiliation{V.N. Karazin Kharkiv National University, Kharkiv, Ukraine}
\author{Georgios Korpas}
\affiliation{HSBC Lab, Emerging Technologies,
Innovation \& Ventures, Singapore}
\affiliation{Faculty of Electrical Engineering,
Czech Technical University in Prague, Czech Republic}
\affiliation{Archimedes Research Unit on AI,
Data Science and Algorithms, Athena Research Center, Greece}
\author{Philippe J.S. De Brouwer}
\affiliation{HSBC Service Delivery Sp.~z o.o., Krakow, Poland}
\author{Mykola Maksymenko}
\affiliation{Haiqu, Inc., 95 Third Street, San Francisco, CA 94103, USA}
\author{Maciej Koch-Janusz}
\affiliation{Haiqu, Inc., 95 Third Street, San Francisco, CA 94103, USA}
\affiliation{Department of Physics, University of Z\"urich, 8057 Z\"urich, Switzerland}

\begin{abstract}
Encoding classical data in a quantum state is a key prerequisite of many quantum algorithms. Recently matrix product state (MPS) methods emerged as the most promising approach for constructing shallow quantum circuits approximating input functions, including probability distributions, with only linear number of gates. We derive rigorous asymptotic expansions for the decay of entanglement across bonds in the MPS representation depending on the smoothness of the input function, real or complex. We also consider the dependence of the entanglement on localization properties and function support. Based on these analytical results we construct an improved MPS-based algorithm yielding shallow and accurate encoding quantum circuits. By using Tensor Cross Interpolation we are able to construct utility-scale quantum circuits in a compute- and memory-efficient way. We validate our methods by loading heavy-tailed distributions, including L\'evy, important in finance, but they apply to any smooth function inputs. We test the performance of the resulting quantum circuits by executing and sampling from them on IBM quantum devices, for up to 156 qubits.
\end{abstract}

\maketitle


\section{Introduction}

\def\thefootnote{*}\footnotetext{These authors contributed equally to this work}

The first step of many quantum algorithms is state preparation, encoding the classical input data in \emph{e.g.~}the amplitudes of a quantum state. For a general input, the depth of the quantum circuit preparing a corresponding quantum state may scale exponentially \cite{shende2005synthesis}, making this key step intractable on real quantum devices. To alleviate this problem different strategies have been proposed, including general methods based on Quantum Generative Adversarial Networks (QGANs)\cite{zoufal2019quantum}, Quantum Fourier Transform (QFT) \cite{garcia2021quantum,PhysRevResearch.2.043442,moosa2023linear}, or handcrafted approaches \cite{dasgupta2022loading}, but all of these have shortcomings in terms of scaling, circuit complexity, trainability or generality, limiting their usefulness in practice. Fortunately, for important classes of data accurate approximate and only linearly deep circuits can be systematically constructed. This is based on two properties: First, the representability of the input data as a matrix product state (MPS)\cite{SCHOLLWOCK_2011, Cirac_2021} [more generally: a tensor network (TN)] of \emph{small bond dimension}. Second, the ability to convert a small bond dimension MPS into a quantum circuit with gate count \emph{linear} in the number of qubits \cite{Ran_2020, Rudolph_2024, Zhou_2021, Ben-Dov_2024, Malz_2024, smith_2024}.

We focus on the general problem of representing 1D real or complex input functions with a quantum circuit. These may represent \emph{e.g.}~the initial state for a quantum partial differential equation (PDE) solver \cite{ljubomir2022quantum}, or a probability density function (PDF). The two key questions we address are how the analytical properties of the input influence the entanglement structure in the intermediate MPS representation, and how this structure can be used to yield even shallower approximating quantum circuits.

For the first question, it is known that while a generic function may require an MPS approximation of an exponential rank (bond dimension), polynomials are represented \emph{exactly} with small bond MPS \cite{Oseledets_2012}. Further, results for ranks and error bounds of low-rank MPS approximations exist  for smooth (infinitely differentiable) functions \cite{Holmes_2020, holmes2_2020, garcia2021quantum, Marin_2023, lindsey_2024}, some of which were used in the previous works on distribution encoding \cite{Gundlapalli_2022,Iaconis_2024, Melnikov_2023, akhalwaya2023modularenginequantummonte,sano_2024,GonzalezConde_2024}. 
Here, we significantly strengthen these analyses by deriving results on the precise structure of the entanglement in the MPS approximation in terms of entanglement spectra and purities. In particular we show universal scaling relations of these entanglement measures for smooth functions, and their breakdown in the absence of smoothness. These allow us to derive analytically functionals controlling \emph{e.g.}~the fidelity of a single layer approximate encoding circuits.

\begin{figure*}[t]    
\includegraphics[width= \linewidth]{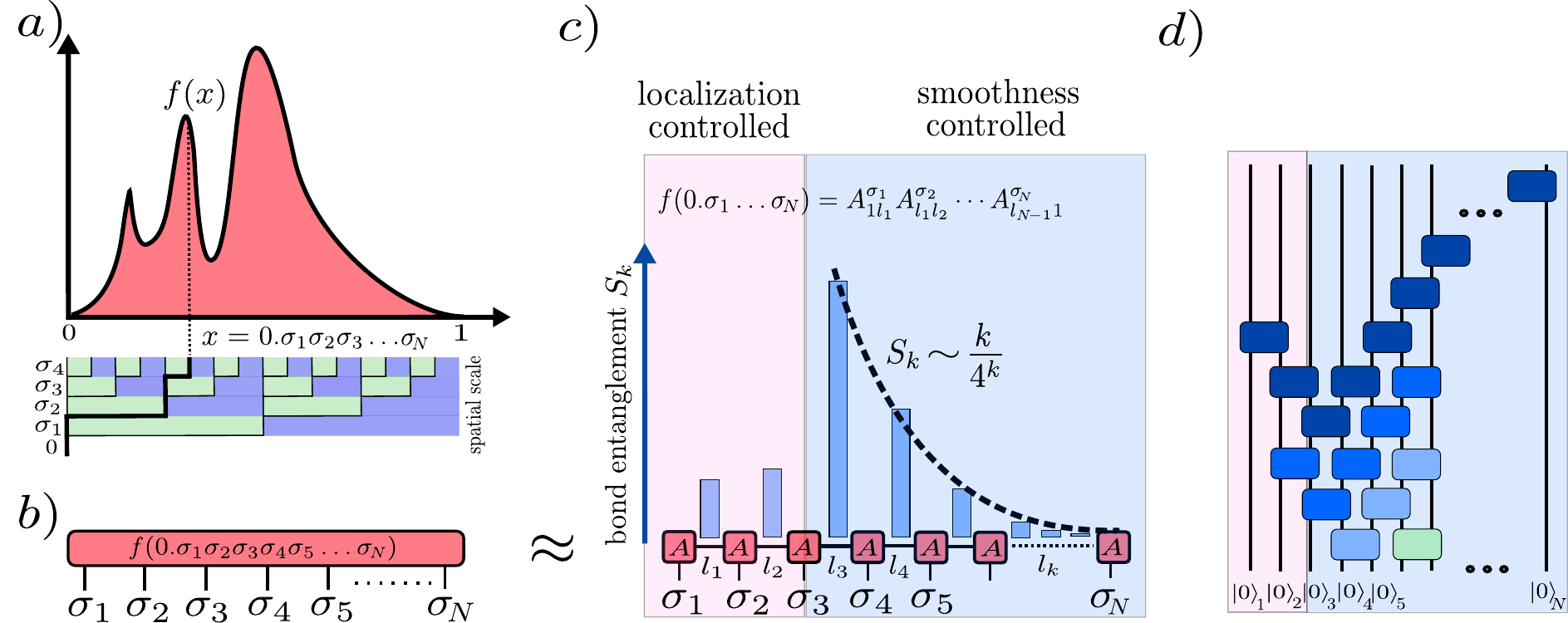}      \caption{\label{fig:mainfig}%
       \textbf{a)} A real or complex function $f(x)$ on $[0,1]$ is discretized on a binary grid $x=\sum_i\sigma_i 2^{-i}$. Successive digits $\sigma_i$ of the binary expansion describe decreasing spatial scales: parts of the interval $[0,1]$ sharing the same value of $\sigma_i$ are shown in green and blue. \textbf{b)} Discretized function can be represented as a rank-$N$ tensor $f_{\sigma_{1}\sigma_{2}\dots\sigma_{N}}$. \textbf{c)} A matrix product state (MPS) can approximate $f_{\sigma_{1}\sigma_{2}\dots\sigma_{N}}$, but the bond dimensions $l_k$, and thus the entanglement across bond $k$, may be exponential in general case. We show that for \emph{smooth functions} the entanglement \emph{decays exponentially} with $k$, effectively decoupling fine spatial scales. For the very largest scales the behavior is non-universal and depends on the localization properties of the function. \textbf{d)} We use these analytical results to construct optimized quantum circuits with $O(N)$ gates generating the state encoding the function $f$. The algorithm minimizes the number of 2-qubit gates based on the entanglement present at different scales.
       }
\end{figure*}

Armed with these results on entanglement generated by approximate function representations, we tackle the second question. We introduce an improved algorithm generating encoding quantum circuits with reduced depth compared to previous MPS-based proposals. We also introduce a number of additional optimizations based on \emph{e.g.~}the compilation of isometries, further reducing the gate count. This is summarised in Fig.~\ref{fig:mainfig}.

Finally, we deploy and test our algorithm in practice on the example of encoding probability distributions. This important case is motivated by their role in quantum finance routines \cite{Woerner_2019_opt_pricing1,Stamatopoulos_2020_opt_pricing2}, particularly of the $\alpha$-stable ones, \emph{i.a.}~the L\'evy distribution, which are used to effectively model the fat tailed or skewed data in risk assessment, portfolio management and derivative pricing \cite{bouchaud1997theorie, cont2004financial,mitocw_levy, mandelbrot1997variation,debrouwer2009maslowian, debrouwer2010target,cont2004financial, mantegna1995econophysics}. We test our methods by executing the circuits on real and simulated IBM quantum devices on up to 156 qubits. In particular we execute and sample from the quantum circuits generated by our method on the \texttt{ibm\_torino} QPU on up to 25 qubits, with the generated distributions passing standard statistical tests (\emph{e.g.}~Kolmogorov-Smirnov). We also execute utility scale circuits for up to 156 qubits on the \texttt{ibm\_marrakesh} and \texttt{ibm\_kingston} QPUs. These we generate using a scalable compute- and memory-efficient Tensor Cross Interpolation \cite{TCI_paper} extension of our algorithm, obtaining qualitative agreement with the exact distributions even for these large utility-scale problems.

The paper is structured as follows: In Sec.\ref{sec:preliminaries} we introduce the problem and the necessary background. In Sec.\ref{sec:bounds} we state our theoretical results, numerical experiments, and discuss their implications. In Sec.\ref{sec:implementation} we give an improved circuit construction algorithm based on them. In Sec.\ref{sec:experiments} we benchmark the results on real and simulated quantum hardware, and in Sec.\ref{sec:conclusions} we discuss future directions. Appendices give mathematical details of the proofs, and further algorithmic details.

\section{Preliminaries: MPS approximation of functions}\label{sec:preliminaries}

\subsection{The MPS representation}

Our goal is to encode a function $f(x) \to \mathbb{C}$ with $x \in [0,1]$, normalized in $L_{2}$ norm: $\int_{0}^{1} |f(x)|^{2} dx = 1$. To this end $f(x)$ is discretized on a $N$ bit binary grid (see Fig.\ref{fig:mainfig}.a):
\begin{equation}
x_{\sigma} = \sum_{i=1}^N \frac{\sigma_i}{2^{i}}  \equiv 0.\sigma_{1}\sigma_{2}\dots\sigma_{N},
\label{eq:bingrid}
\end{equation}
where $\sigma = (\sigma_1,\dots,\sigma_N)$ with $\sigma_i\in\{0,1\}$.
Evaluating $f(x)$ on this grid yields a rank-$N$ tensor $f_{\sigma_{1}\sigma_{2}\dots\sigma_{N}}$, with a  $O(1/2^N)$ discretization error. These $2^N$ values will be (approximately) encoded, using a quantum circuit, in an $N$-qubit quantum state $|f\rangle$ (see Fig.~\ref{fig:mainfig}b):
\begin{equation}
 |f\rangle =  \sum_{\sigma_1,\ldots,\sigma_N = 0}^1 \frac{f_{\sigma_{1}\sigma_{2}\dots\sigma_{N}}}{\sqrt{2^N}}|\sigma_{1}\sigma_{2}\dots\sigma_{N}\rangle, 
 \label{eq:ftensor}
\end{equation}
where $\sqrt{2^{N}}$ ensures the proper normalization. Note that this encoding is hierarchical: the largest indices (qubits) $\ldots,\sigma_{N-1}, \sigma_{N}$ describe the function on the smallest scales of order $1/2^{N}$, while the smallest indices $\sigma_{1}, \sigma_{2},\ldots$ account for the largest scales.

The exponentially sized tensor $f$ can be rewritten and decomposed into a matrix product state (MPS) \cite{SCHOLLWOCK_2011, Cirac_2021}, also known as quantized tensor train \cite{khoromskij_2011}, (see Fig.\ref{fig:mainfig}c): 
\begin{equation}\label{eq:tensor_f_approx}
    |f\rangle \approx \sum_{\sigma_1,\ldots,\sigma_N = 0}^1 c_{\sigma_{1}\sigma_{2}\dots\sigma_{N}}|\sigma_{1}\sigma_{2}\dots\sigma_{N}\rangle,
\end{equation}
where
\begin{equation}\label{eq:noncanonical}             c_{\sigma_{1}\sigma_{2}\dots\sigma_{N}} = A_{1,l_{1}}^{\sigma_{1}} A_{2,l_{1} l_{2}}^{\sigma_{2}} \dots A_{N-1, l_{N-2}l_{N-1}}^{\sigma_{N-1}} A_{N, l_{N-1}}^{\sigma_{N}},
\end{equation}
and where $A^{\sigma_k}_{k,l_{k-1}l_k}$ are rank-$3$ tensors with a ``physical" index $\sigma_k$, and ``internal" indices $l_{k-1}$,$l_k$. The summation range of the contracted internal indices (we use the Einstein summation convention), also known as bonds, is called the bond dimension $\chi_k$. Equivalently, the above tensor decomposition can be written in the so-called ``canonical" form as:
\begin{align}\label{eq:canonical}                  & c_{\sigma_{1}\sigma_{2}\dots\sigma_{N}} = \\& \nonumber \Gamma^{\sigma_{1}}_{1, l_{1}} \Lambda_{1, l_{1}} \Gamma ^{\sigma_{2}}_{2, l_{1} l_{2}} \Lambda_{2, l_{2}}\dots \Gamma^{\sigma_{N-1}}_{N-1, l_{N-2}, l_{N-1}} \Lambda_{N-1, l_{N-1}} \Gamma^{\sigma_{N}}_{N, l_{N-1}},
\end{align}
where $\Gamma_{k}$ are also rank-$3$ tensors, while $\Lambda_{k}$ are positive diagonal matrices, which obey certain isometrical conditions (for a thorough introduction to MPS see \cite{SCHOLLWOCK_2011,Vidal_2003,perezgarcia2007matrixproductstaterepresentations_mps,Xiang_2023_mps_book}).
 These canonical form conditions have the form:

\begin{align}\label{right_isometry}
    \sum_{l_{k}, \sigma_{k}}\Gamma_{k, l_{k-1}, l_{k}}^{\sigma_{k}} \Lambda^{2}_{k, l_{k}} \Gamma_{k, l_{k-1}', l_{k}}^{\dagger \sigma_{k}} &= \delta_{l_{k-1}, l_{k-1}'}, \\
\label{left_isometry}
    \sum_{l_{k-1}, \sigma_{k}}\Gamma_{k, l_{k-1}, l_{k}}^{\sigma_{k}} \Lambda^{2}_{k-1, l_{k-1}} \Gamma_{k, l_{k-1}, l_{k}'}^{\dagger \sigma_{k}} &= \delta_{l_{k}, l_{k}'}.
\end{align}

Importantly, the entries of the diagonal matrices $\Lambda$ in the canonical form encode the MPS entanglement spectra. To wit, the entanglement entropy for the bipartition on the bond $k$ is given by: 

\begin{equation}
    S_k = -\sum_{i=0}^{\chi_{k}-1}\Lambda_{k, i}^{2} \log{\Lambda_{k, i}^{2}},
\end{equation}
where the summation range over virtual bond dimension index $i$ starts from $0$. 
We will also use a different entanglement measure, the purity, which is easier to compute for an arbitrary function. The purity on bond $k$ is given by $p_{k} = \mathrm{Tr} \rho_{k}^{2}$, where $\rho_{k}$ is the reduced density matrix for the subsystem $\sigma_{1} \sigma_{2} \dots \sigma_{k}$. It characterizes how close the state is to the product state. In particular, if the initial wave function factorizes with respect to the bond $k$, then the purity is equal to one. The purities $p_k$ are given by:


\begin{equation}
    p_{k} = \sum_{i=0}^{\chi_{k}-1} \Lambda_{k, i}^{4}.
    \label{eq:purspect}
\end{equation}

In what follows the MPS approximation of the function will be converted to a shallow quantum circuit preparing the encoded function on a QPU (Fig.\ref{fig:mainfig}d) using our improved algorithm (see Sec.\ref{sec:implementation}), which extends the methods of Refs.\cite{Gundlapalli_2022,Iaconis_2024, Melnikov_2023, sano_2024,GonzalezConde_2024} (see also \cite{PhysRevA.101.010301}).

We will also find useful in Sec.\ref{sec:implementation} the center canonical form (also known as mixed canonical form, see Ref.~\cite{SCHOLLWOCK_2011}). In this form, some bond (and the corresponding $\Lambda$ matrix) is chosen as central. Then, all the $\Lambda$ matrices on the other bonds are absorbed into their neighbouring $\Gamma$ tensors, so that resulting combined tensors become isometries. The MPS  thus obtained (here with the $k$th bond being central) has the form: 

\begin{equation}\label{eq:center_canonical}            A_{L, 1,l_{1}}^{\sigma_{1}}  \dots A_{L,k, l_{k-1}l_{k}}^{\sigma_{k}}\Lambda_{k, l_{k}} A_{R,k+1,l_{k}l_{k+1}}^{\sigma_{k+1}}\dots  A_{R, N, l_{N-1}}^{\sigma_{N}},
\end{equation}
where the tensors $A_{L}$ and $A_{R}$ obey the left isometrical and right isometrical conditions, respectively: 

\begin{align}\label{right_canonical}
    \sum_{l_{k}, \sigma_{k}}A_{R,k, l_{k-1}, l_{k}}^{\sigma_{k}}  A_{R, k, l_{k-1}', l_{k}}^{\dagger \sigma_{k}} &= \delta_{l_{k-1}, l_{k-1}'},\\
\label{left_canonical}
    \sum_{l_{k-1}, \sigma_{k}}A_{L, k, l_{k-1}, l_{k}}^{\sigma_{k}} A_{L, k, l_{k-1}, l_{k}'}^{\dagger \sigma_{k}} &= \delta_{l_{k}, l_{k}'}.
\end{align}

The position of the center in the center canonical form can be shifted by absorbing the central bond $\Lambda$ matrix into the neighbouring $A_{L}$ or $A_{R}$ tensors and proceeding with SVD of the resulting tensor. To wit, to shift the center position from bond $k$ to bond $k-1$, we first define a new tensor $A_{C, k, l_{k-1}l_{k}}^{\sigma_{k}} = A_{L,k, l_{k-1}l_{k}}^{\sigma_{k}}\Lambda_{k, l_{k}}$, which we then decompose with SVD: $A_{C, k, l_{k-1}l_{k}}^{\sigma_{k}} = U_{l_{k-1}l_{k-1}'} \Lambda_{k-1, l_{k-1}'}A_{R, k, l_{k-1}' l_{k}}^{\sigma_{k}}$, where $\Lambda$ is the new central bond $\Lambda$ matrix, $A_{R, k}$ is the new right-isometric tensor on the $k$-th qubit and $l_{k-1}'$ is the new index on $(k-1)$-th bond. Finally, the matrix $U_{l_{k-1} l_{k-1}'}$ is absorbed into the tensor $A_{L, k-1}$: $A_{L, k-1, l_{k-2} l_{k-1}}^{\sigma_{k-1}} \to A_{L, k-1, l_{k-2} l_{k-1}'}^{\sigma_{k-1}} = A_{L, k-1, l_{k-2} l_{k-1}}^{\sigma_{k-1}} U_{l_{k-1} l_{k-1}'}$.  

\subsection{Constructing the MPS}

While the MPS decomposition can be exact, for arbitrary tensors the MPS ranks (bond dimensions) may be exponential in the number of qubits. Fortunately, in many important cases this is not the case, as \emph{e.g.}~when representing the ground states of quantum Hamiltonians \cite{SCHOLLWOCK_2011}. Thus if the entanglement in $|f\rangle$ in Eq.\eqref{eq:ftensor} is not large, the state can be effectively MPS approximated with polynomial resources by truncating to small bond dimensions in Eqs.\ref{eq:tensor_f_approx},\ref{eq:noncanonical},\ref{eq:canonical}. Later we show that this holds for smooth functions. 

The MPS decomposition itself is carried out either with singular value decomposition (SVD) or approximate methods ~\cite{lindsey_2024, rodríguezaldavero_2024, Oseledets_2010, Ritter_2024, 6076873_oracle}. The SVD-based method works as follows: the tensor $c_{\sigma_{1}, \sigma_{2} \dots \sigma_{N}}$ is interpreted as a matrix $c^{\sigma_1\rho}$ whose first index is $\sigma_{1}$, and the second one comprises all of the remaining ones: $\rho = (\sigma_2\ldots\sigma_n)$. SVD decomposes this matrix into the form: 
\begin{align}
c^{\sigma_1\rho} = A^{\sigma_{1}}_{1,l_{1}} s_{l_{1}}V_{l_{1}}^\rho &= A^{\sigma_{1}}_{1,l_{1}} \Lambda_{1, l_{1}}V_{l_{1}}^{\sigma_{2} \sigma_{3}\dots\sigma_{N}} \nonumber 
\end{align}
The tensor $\Lambda_{1, l_{1}}V_{l_{1}}^{\sigma_{2} \sigma_{3}\dots\sigma_{N}} \nonumber $ can again be interpreted as a matrix with indices $\sigma_2$ and $(\sigma_3\ldots\sigma_N)$ and be decomposed with SVD once more: $\Lambda_{1, l_{1}}V_{l_{1}}^{\sigma_{2} \sigma_{3}\dots\sigma_{N}} \nonumber  = A^{\sigma_{2}}_{2, l_{1} l_{2}} \Lambda_{2, l_{2}} \tilde{V}^{\sigma_{3}\dots\sigma_{N}}_{l_{2}}$. This is iterated until the original tensor $c$ is fully factorized into the product of tensors $A^{\sigma_{k}}_{k, l_{k-1} l_{k}}$ and a residual tensor $A^{\sigma_{N}}_{N, l_{N-1}} = \Lambda_{N-1, l_{N-1}}V^{\sigma_{N}}_{l_{N-1}}$. This factorized form can be exact, but in an MPS approximation SVD spectra $\Lambda_{l}$ are truncated below a small threshold, which we set to $10^{-12}$. This is how Eq.~\ref{eq:noncanonical} is obtained.

The MPS in a canonical form Eq.~\ref{eq:canonical} is obtained by inserting $1 = \Lambda_{\alpha} \cdot \Lambda_{\alpha}^{-1}$ on the bonds, and contracting the inverse $\Lambda_{\alpha}^{-1}$ with the following $A$ tensor, obtaining the $\Gamma$ tensors. Due to the SVD spectra truncation the canonical conditions Eqs.\ref{right_isometry},\ref{left_isometry} may not be satisfied exactly. To correct this one more MPS sweep is performed.

The main limitation of the SVD method is that the original tensor $c_{\sigma_{1} \sigma_{2} \dots \sigma_{N}}$ has to be stored explicitly, implying a $O(2^N)$ memory scaling. This can only be done for moderate numbers of qubits ($\sim 30$). For larger problems the MPS has to be constructed differently. We use tensor cross interpolation (TCI)~\cite{TCI_paper,6076873_oracle}, which allows an MPS approximation to be found provided only access to a callable function (\emph{i.e.}~the ability to evaluate the function at different points) without requiring to store the complete set of function values. 
TCI constructs the MPS representation of the function by sampling of the function's values, which it uses for interpolation. As these are the only values stored, it drastically reduces the memory requirements, allowing to construct MPS representation for large qubit numbers. 
The number of function evaluations in TCI scales with the number of qubits as $O(N\chi^2)$, where $\chi$ is the maximal MPS bond dimension. This is in contrast to the standard SVD approach which involves the calculation of a whole state vector with $2^N$ function evaluations. As its name implies, TCI is a generalisation of matrix cross-interpolation \cite{bebendorf2000approximation}, which constructs a low-rank approximation of a general matrix from a sparse set of its rows and columns. In TCI this is applied iteratively bond by bond, as in the SVD case.

In what follows we will see that smooth functions do not require large MPS bonds to be represented accurately. One further advantage of the TCI method is that it does not require explicit representation of the of the function being sampled; it can be provided as a black box. For a pedagogical introduction and technical details of the TCI algorithm we refer the reader to Refs.~\cite{TCI_paper,6076873_oracle}.

\section{Entanglement in tensorial approximants of smooth functions}\label{sec:bounds}

\subsection{General results}\label{sec:generalres}

Here we summarize our theoretical results on the behavior of entanglement in tensor approximations of smooth functions. Particularly, we derive universal scaling relations for purities and entanglement spectra for the qubits describing the smallest scales. For the qubits describing the largest scales, outside the universal regime, we note the dependence of the entanglement spectra on function concentration. Taken together, these results enable valuable insights into the construction and precision of the quantum circuit encoding of the function. Proofs of the theorems and corollaries are provided in Appendix~\ref{app:A}.

\begin{theorem}\label{thm:purities}
    For a smooth (having infinitely many continuous derivatives) and normalized function $f(x)$ on the interval [0,1] the purities $p_{k}$ of the tensorial representation $f_{\sigma_{1}\sigma_{2}\dots\sigma_{N}}$ decay exponentially with the bond number $k$ (\emph{i.e.}~with decreasing spatial scale of function variations, via Eq.\eqref{eq:bingrid}). More precisely, as $k\to\infty$ we have:
    \begin{equation}\label{eq: purity}
    p_{k} = 1 - g_1(f)/(6 \times 4^{k}) + O(1/8^{k}),
    \end{equation}
    where:
    \begin{align}
        g_1(f) = & \int_0^1 f'(u) (f^*)'(u) du + \nonumber \\ 
        & - \int_0^1 f'(u) f^*(u) du \times \int_0^1 f(u) (f^*)'(u) du, \label{eq:g1f}
    \end{align}
    and where $f'$ and $f^*$ denote the derivative and complex conjugate of $f$, respectively.
\end{theorem}
\noindent

In Appendix~\ref{app:A} we show that $g_1(f)$ vanishes precisely for functions which are expressible as trivially factorizable MPS of bond dimension one. 
A key consequence of Theorem~\ref{thm:purities} is that a state encoding a smooth function approaches a factorized state with increasing $k$.

We further study the reduced density matrices obtained from a bipartition of the qubit system. 
Let us thus consider the reduced density matrix of the first $k$ qubits $\rho_k$. We have:
\begin{theorem}\label{thm:density_matrix}
Under the same assumptions as in Theorem 1, to leading order in $1/2^k$, the eigenvalues of the reduced density matrix  of the function encoding on the first $k$ qubits scale as $\rho_{k,n} = O(1/4^{kn})$, where $n\in \mathbb{N}_0$ corresponds to $n$-th eigenvalue and $k\to\infty$. 
\end{theorem}

\begin{figure*}[t] 
\centering
\includegraphics[width=0.85\linewidth]{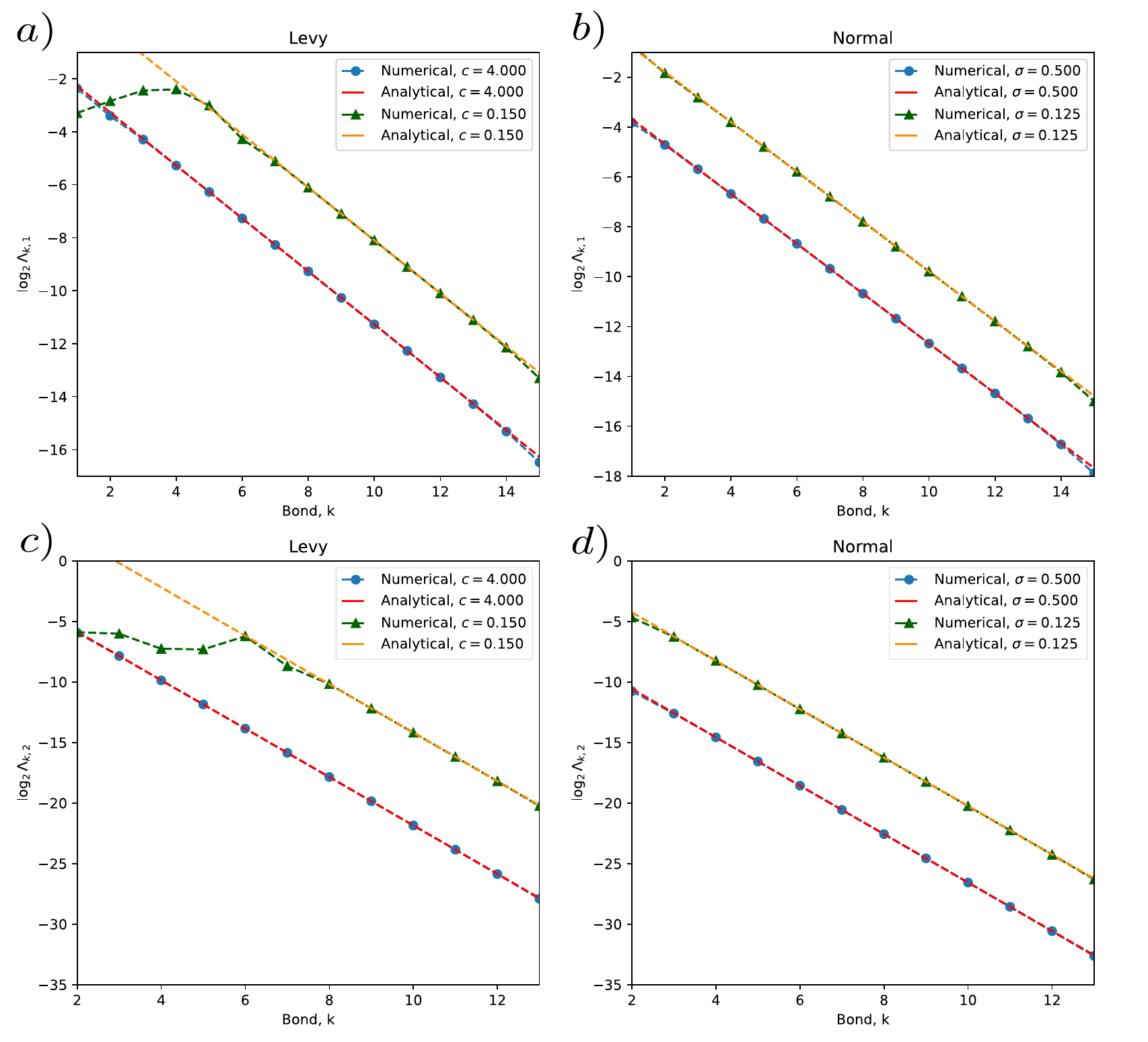} 
     \caption{Comparison of the analytical (see Theorem~\ref{thm:purities} and Corollary~\ref{cor:eigenvalue}) and numerical SVD results on the dependence of the entanglement spectra $\Lambda_{k,1}$ and $\Lambda_{k,2}$ on bond position $k$. \textbf{a)} $\Lambda_{k,1}$ for the L\'evy distribution with $c = 0.15$ and $c=4$. \textbf{b)} $\Lambda_{k,1}$ for the normal distribution with $\mu = 0.5$ for $\sigma = 0.125$ and $\sigma =0.5$. \textbf{c)} $\Lambda_{k,2}$ for the L\'evy distribution with $c = 0.15$ and $c = 4$. \textbf{d)} $\Lambda_{k,2}$ for the normal distribution with $\mu = 0.5$ for $\sigma = 0.125$ and $\sigma =0.5$. Note the excellent agreement and universal behaviour for large $k$, and non-universal features for small $k$, as discussed in the text.}
     \label{fig:spectra1}
\end{figure*}

Further, from the above results on the scaling of purities and eigenvalues we immediately obtain corollaries about the behavior of entanglement spectra and entanglement entropy. To wit, the relation between purities and entanglement spectra is given by Eq.\eqref{eq:purspect}. As the purity approaches unity, the leading coefficient $\Lambda_{k,0}$ dominates and others vanish. More specifically:
\begin{corollary}\label{cor:eigenvalue}
For a smooth and normalized on $[0,1]$ function $f(x)$ the subleading coefficient $\Lambda_{k,1}$ scales as:
\begin{equation}\label{eq:lambda1k}
\Lambda_{k,1} \sim \frac{1}{2^k}\sqrt{\frac{g_{1}(f)}{12}}
\end{equation}
as $k\to\infty$, where $g_1$ is defined in Theorem~\ref{thm:purities}.
The leading entanglement spectrum coefficient scales as
$\Lambda_{k,0} \sim 1 - g_{1}(f)/(24 \cdot 4^{k})$
\end{corollary}

\begin{corollary}\label{cor:entropy}
 The entanglement entropy $S_k$ decays exponentially with the bond index $k$:
 \begin{equation}
     S_k = O\left(\frac{k}{4^k}\right) \mbox{\ \ \  as \ \ \ } k\to\infty 
 \end{equation}
\end{corollary}

In particular, from Theorem~\ref{thm:density_matrix} for the eigenvalue $\rho_{k,1}$ we obtain the asymptotic $\rho_{k,1} \sim g_{1}(f) / (12 \times 4^{k})$, which agrees with  Corollary~\ref{cor:eigenvalue}. The next eigenvalue $\rho_{k,2}$ can also be relatively simply computed, which we do in Appendix~\ref{app:B}. We will also use this value in the further numerical analysis. In Appendix~\ref{app:B} we also provide several analytical examples to illustrate this scaling.
Additionally, in Appendix \ref{subsec_smoothness} we observe that if the function is not smooth, but has absolutely continuous derivatives up to $r$-th order, then our previous results hold for entanglement spectra up to order $r+1$, while the $r+2$ and higher orders are non-universal.

The above results are based on function smoothness, and characterize the asymptotic behavior for qubits describing the smallest spatial scales of function's variations (as per Eq.\ref{eq:bingrid}). The entanglement spectra for the leading qubits, corresponding to the large scales, are non-universal, but we observe a strong dependence on the localization properties of the function. For example, a maximally localized delta-peaked distribution is described by an entirely unentangled product state. Similarly, as described in detail below, exponentially decaying functions are less entangled than the power-law tailed ones.

Let us compare our results with those previously obtained in the literature. In Ref.~\cite{garcia2021quantum} a bound for the purity was obtained in the following form: $p_{k} \geq 1 - 4 \max |f(x)f'(x)|/2^{k}$ for a positive $f$. In contrast, our analytic expansions for purities show that the first non-vanishing term scales $1/4^{k}$ for a general $f$, \emph{i.e.}~an exponential improvement in the bound.
In Ref.\cite{Holmes_2020} an Ansatz for entanglement spectra of the form $\sqrt{\rho_{kn}} = \Lambda_{k,n} = \alpha_{k} \exp{(-\beta_{k} n)}$ is used, and is fit to empirical data. While this captures the exponential dependence on $n$, and is thus consistent with our results, we emphasize that in our case this scaling dependence is analytically derived, and not empirically postulated. Finally, Ref.\cite{lindsey_2024} studied approximability of functions with matrix product states. Assuming extendibility to the complex plane they bounded the maximal deviation between the function and its approximation with MPS with $k$-th bond dimension truncated to $\chi$ by $O(2^{-k\chi})$. While Ref.\cite{lindsey_2024} does not study the entanglement properties of the encoding, this result agrees with our predictions, as the maximal deviation should be of the same order as the maximal truncated entanglement spectra, which in this case is precisely of the order $O(2^{-k\chi})$. Moreover, our results allow to compute the exact coefficients in the obtained expansions and bounds, which we demonstrate for probability distributions.

\begin{figure*}  
\centering\includegraphics[width= 0.9\linewidth]{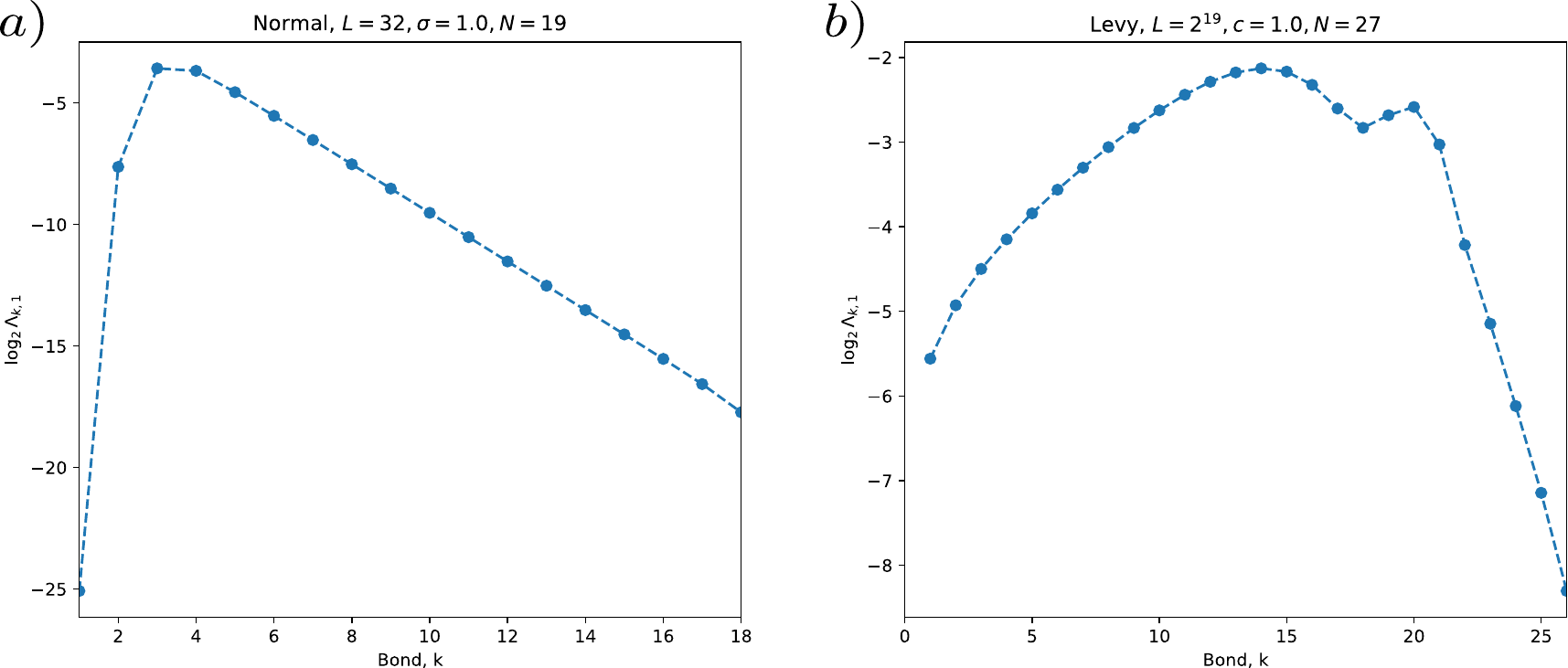}      \caption{\label{fig2: localization}%
       Entanglement decay for exponential and polynomial localization: \textbf{a)} The entanglement spectra $\Lambda_{k,1}$ as a function of the bond position $k$ (lines are for guidance only) for the exponentially localized one-sided normal ($\mu=0$, $\sigma = 1$) distribution truncated to the interval $[0, 32]$, for $N=19$ qubits. Note the super-exponential decay of entanglement for small $k$, and the universal, exponential, smoothness-controlled regime at large $k$. \textbf{b)} The entanglement spectra $\Lambda_{k,1}$ as a function of the bond position $k$ for the L\'evy distribution (with $c=1$ and interval $[0, 2^{19}]$), on $N=27$ qubits. For small $k$ the entanglement decays super-exponentially, but, due to only polynomial localization, slower than for normal distribution, delaying the onset of the universal regime at large $k$.
       }
\end{figure*}

\subsection{Encoding probability distributions}

While our results hold for a more general classes of functions (\emph{e.g.}~complex valued), we now focus on the important problem of encoding probability distributions $p(x)$. This is special case of function encoding with $f(x) = \sqrt{p(x)}$, where the normalization of the probability corresponds to $L_2$ normalization of $f$. These functions, particularly modeling heavy-tailed distributions, have practical applications in areas such as portfolio optimization and derivative pricing \cite{cont2004financial,carr2002fine}, and risk management \cite{cont2004financial, bouchaud1997theorie}.

We consider the following examples of smooth distributions on the interval $[0,1]$: the (truncated) normal, log-normal and L\'evy distributions:
\begin{align}
    p(x)_{normal} = \frac{1}{\sqrt{2 \pi \sigma^{2}}} \exp{\left[-\frac{(x - \mu)^{2}}{2\sigma^{2}}\right]}, \\
    p(x)_{log-normal} = \frac{1}{x\sqrt{2 \pi \sigma^{2}}} \exp{\left[-\frac{(\log{x} - \mu)^{2}}{2\sigma^{2}}\right]}, \\
    p(x)_{Levy} = \sqrt{\frac{c}{2 \pi}} \frac{\exp{\left[-\frac{c}{2 x}\right]}}{x^{3/2}}.
\end{align}
We generally set location parameters $\mu$ to $1/2$ so the distributions are mostly located in $[0,1]$ interval, and vary the scale parameters $\sigma$ or $c$. We note that an additional (often negligible) normalization is required due to truncation of the support to a finite interval. The functionals $g_{1}(f)$ from Eq.\eqref{eq:g1f}, as well as quantities related to the third eigenvalue $\Lambda_{k,2}$, used below, are computed explicitly in Appendix~\ref{app:B}.

In Fig.\ref{fig:spectra1}a we compare the subleading entanglement eigenvalue, which, via Eq.\eqref{eq:lambda1k} is given by $\Lambda_{k,1} \sim \sqrt{g_{1}(f)}$, for the L\'evy distribution truncated to the interval $[0,1]$, to the values obtained from performing SVD on the input tensor $f_{\sigma_{1}\sigma_{2}\dots\sigma_{N}}$ of the discretized function values.
For large $k$ we find the precise agreement between the analytical and numerical results, as expected from Theorem~\ref{thm:purities} and Corollary~\ref{cor:eigenvalue}, while for the leading bonds, \emph{i.e.~}large scales, the behavior of entanglement is non-universal, as discussed before.

In Fig.\ref{fig:spectra1}b we compare the same property for the normal distribution (with $\sigma = 1/2$ and $\sigma = 1/8$). We obtain a nearly perfect match between the analytical prediction and numerics. The small discrepancy on the last bond is due to the truncation of the function's discretization to the finite scale $1/2^{N}$, while the discrepancy on the first bonds is caused by the subleading corrections in $1/2^{k}$. 

Finally, in Figs.\ref{fig:spectra1}c,d we compute and compare the third entanglement eigenvalue $\Lambda_{k,2}$ for L\'evy and normal distributions, respectively. As expected, our analytical result matches the numerical data, which indeed behave as $O(1/4^{k})$. As in the previous cases, a discrepancy with analytic results appears only on the first bonds corresponding to the large scales.

\begin{figure*}[t]    \includegraphics[width = \linewidth]{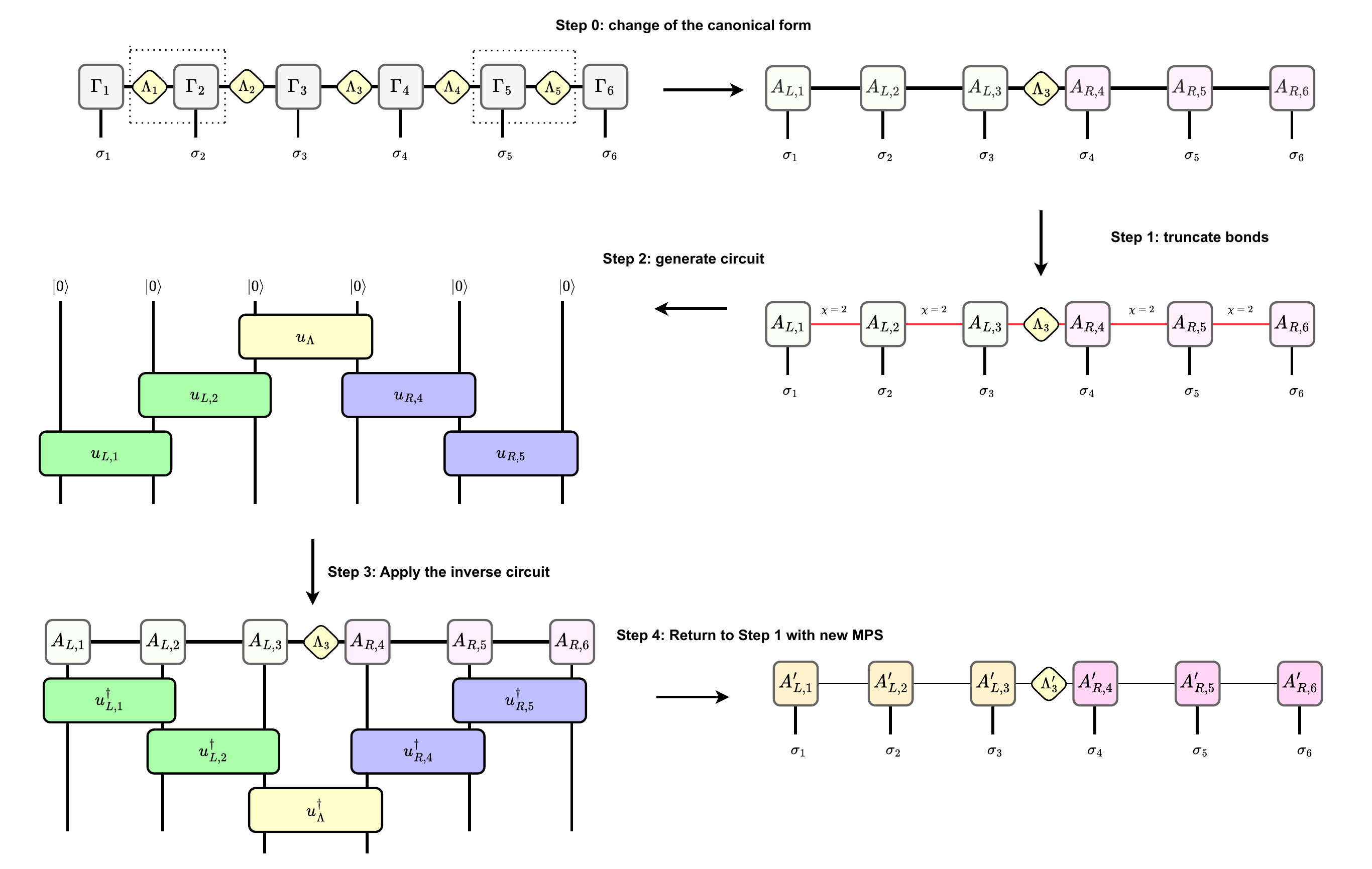}      \caption{\label{fig: analytic_decomposition}%
       An overview of our MPS based-circuit construction algorithm. First, the standard canonical form MPS is transformed into the central canonical form (Step 0). All of the bonds are then truncated to bond dimension $\chi = 2$, shown in red (Step 1). The truncated MPS is decomposed exactly into a unitary circuit using V-layers [see the main text], which may also be incomplete. The central gate may be variationally optimized. (Step 2). On an MPS simulator the inverse of the circuit is appended to the original non-truncated MPS to serve as a disentangler (Step 3). A new MPS is obtained by contracting the tensor network, which is now less entangled. The procedure is repeated for the new state.
       }
\end{figure*}

\subsection{Encoding dependence on the localization and support of the function}

Previously we studied the entanglement on the smallest scales and showed it exhibits a universal exponential decay. While no such universality holds true for the largest scales (see Appendix \ref{app:C} for details) we can nevertheless understand the entanglement at large scales better for a specific class of functions, namely ones exhibiting \emph{localization}. This class contains in particular the probability distributions we were concerned with in the preceding section.

For this discussion it will be useful to bring back the explicit dependence on the support size. Consider a function $f(x)$ defined on $[0, \infty)$. Numerically, to represent the function its support will be truncated to a chosen finite interval $[0,L]$, the function then normalized in $L_2$ norm, and discretized at a fixed resolution $L/2^{N}$ (using $N$ qubits). This is of course equivalent to encoding $\sqrt{L} f(Lx)$, with $x \in [0,1]$. To iteratively increase the support size at a fixed resolution we may simply double the interval to $[0,2L]$ while simultaneously adding another qubit (adjusting the normalization, as required).

 We now assume that $f(x)$ is localized, exponentially or polynomially, without loss of generality on the sub-interval $[0,L]$ of $[0, \infty)$. By this we mean that the mass is concentrated on a finite interval $[0,L]$ and is vanishing in the distribution tail. More formally, for all $n$ the integral of $|f|^2$ on the subinterval $[0, 2^{n-1}L]$ is close to $1$, while it is exponentially (resp.~polynomially) small in the subinterval size on $[2^{n-1} L, 2^{n} L]$ as we increase $n$.

 We can now examine the entanglement in the highest scale qubits, as we successively increase the support size at fixed resolution, using the doubling procedure described above.
 As discussed in more detail in Appendix \ref{app:C} the $|f|^2$ integrals over $[0,2^{n-1}L]$ are in fact equal to the matrix elements of the reduced density matrix obtained by tracing out all but the highest scale qubit. The off-diagonal elements, and thus also the entanglement, can easily be estimated to be of order $\exp(-\lambda 2^{n-1}L)$ for some $\lambda>0$ in the exponential localization case, and of order $(2^{n-1}L)^{-\alpha+1}$ in the polynomial case, where $\lambda$ and $\alpha$ characterize the exponential/polynomial tail. Thus, as we add qubits to represent the increasing spatial scales (and increase the size of the support to which the function is truncated), the entanglement of the leading qubits with those qubits representing lower scales decreases super-exponentially in $n$ for a exponentially localized function, and at least exponentially in $n$ for a polynomially localized function. For large enough $n$ the qubits representing the large scales of the localized function will be almost fully disentangled from the qubits representing the lowest scales. This argument allows us to understand how large an interval is needed to faithfully represent a function.

\begin{figure}[t]
\centering
\includegraphics[width= \columnwidth]{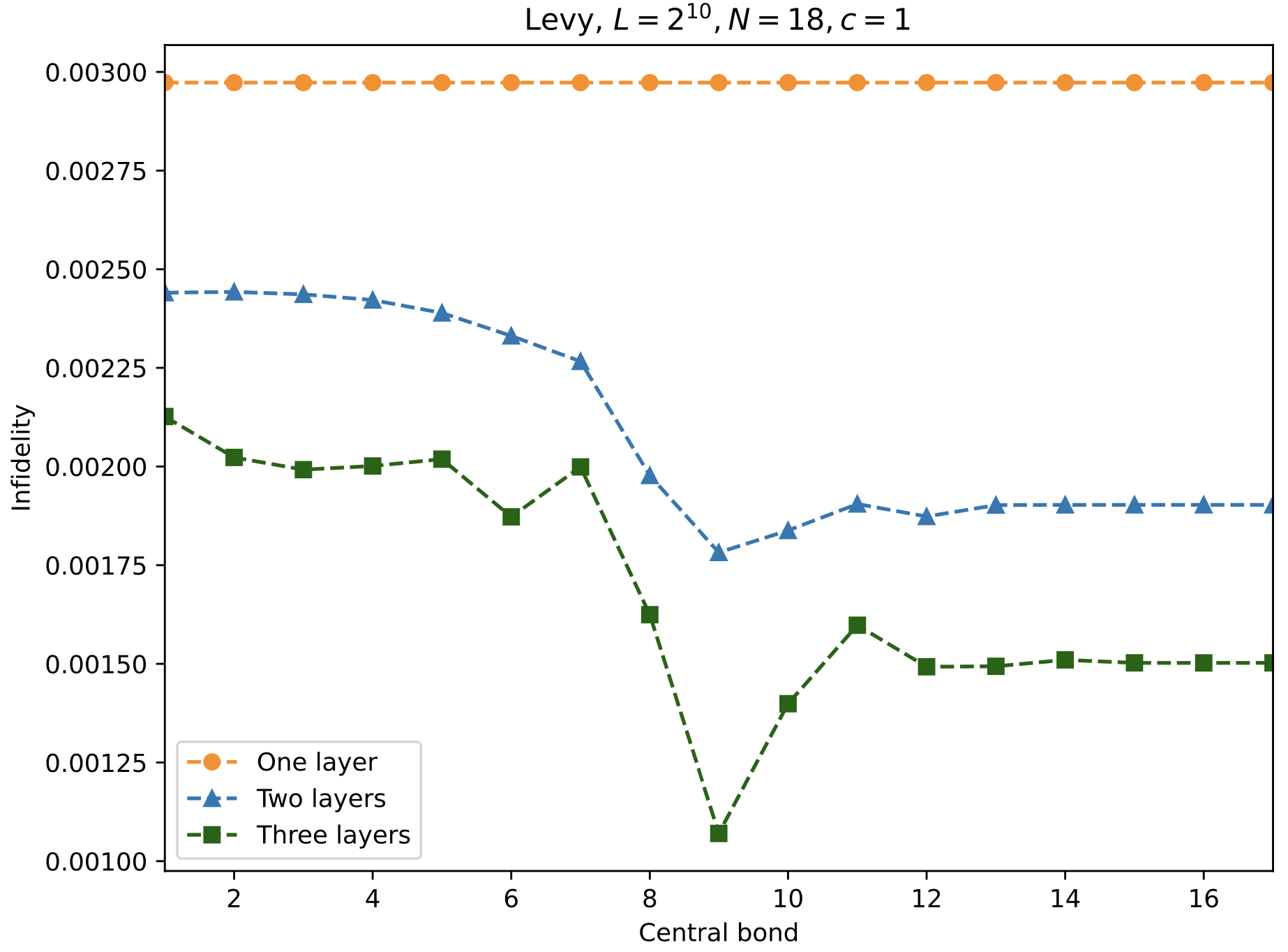}      \caption{\label{fig: Levy_centre_position2}%
      The dependence of the infidelity on the number and position of the origin of the V-layers in the encoding quantum circuits on an ideal simulator (lines between points for guidance only). Here we show data for the L\'evy distribution with $c=1$ on support truncated to $L = 2^{10}$ using $N=18$ qubits. 
      The accuracy increases with additional disentangling layers (but see Fig.\ref{fig:Levy_scaled_noise} for the effects of noise), and significantly depends on the origin of the V-layers. The effect of the latter, which also reduces to overall circuit depth, has effect comparable to adding a new disentangling layer.
       }
\end{figure}

The above argument is borne out by numerical experiments.
In Figs.\ref{fig2: localization}a,b we plot the logarithm of the second entanglement eigenvalue $\Lambda_{k,1}$ on bond $k$ between qubits $k$ and $k+1$, for the normal and L\'evy distributions, respectively. The large-$k$ tails are universal and exponentially decaying, as derived in Sec.\ref{sec:generalres}. The low-$k$ behaviour, however, demonstrates a non-universal decay of entanglement towards large scales. Moreover, there is a clear distinction between the exponential localization of the normal distribution and the weaker polynomial one of the L\'evy distribution. For the former, in Fig.\ref{fig2: localization}a, the large scale (small $k$) drop of entanglement is almost instantaneous (superexponential), and also the universal asymptotic regime of exponential decay for lower scales (large $k$) is visible already for a support truncated to a small interval $[0, 2^4]$. Note that the maximum is achieved between $k=3,4$, consistent with the condition $k \approx \log_{2}{(L/2\sigma)}$ derived in Sec.\ref{sec:generalres}. For the L\'evy case the small-$k$ decay of entanglement is much slower (nearly exponential, as expected for polynomial localization), and correspondingly the onset of the universal regime of entanglement decay at small scales is pushed to much larger values of $k$: we required a support truncated to a very large interval $[0, 2^{19}]$ in Fig.\ref{fig2: localization}b to clearly demonstrate both the low- and high-$k$ regimes.

Both Figs.\ref{fig2: localization}a,b demonstrate the characteristic triangular shape, which is also seen for other localized distributions (\emph{e.g.}~log-normal, Gamma, Cauchy). We emphasize that while the large-$k$ behaviour is universal, the small-$k$ is not, and depends, among others, on the detailed localization properties of the function. The slow decay of entanglement for heavy tailed-distributions (\emph{e.g.}~all $\alpha$-stable distributions except normal) has practical implications for the design of encoding circuits, as it indicates more qubits with significant entanglement will be needed, before the onset of the universal exponential decay regime controlled by smoothness.

\begin{figure*}[t]
\centering
\includegraphics[width= 0.85\textwidth]{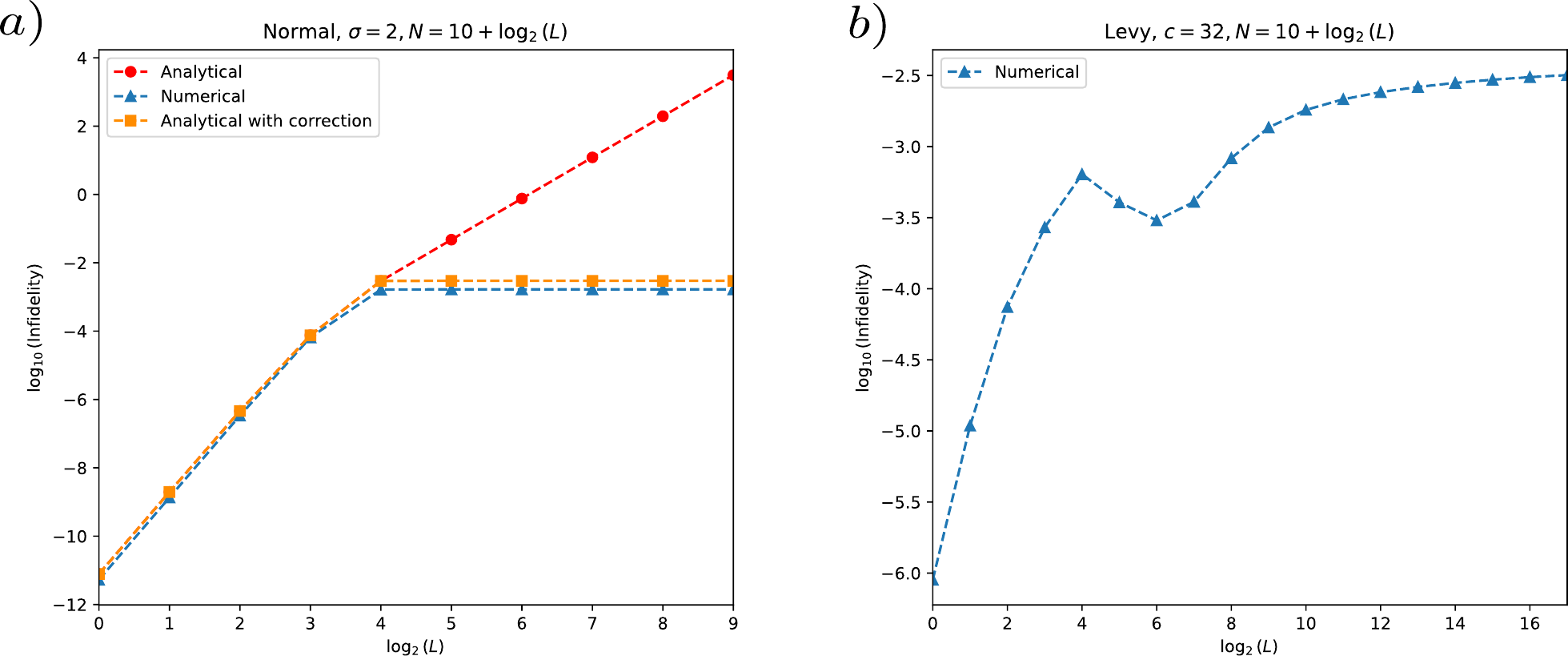}      \caption{\label{fig: normal_error}
      \textbf{a)} The log-log plot of infidelity of a 1-layer quantum circuit as a function of the size of the truncated support interval $L$. We fix the discretization step size, with the qubit number given by $N=10+\lceil\log_{2}L\rceil$. Here we consider the truncated normal distribution with $\sigma =2$, show the numerically computed infidelity, and compare it to the analytical estimates. The label ``Analytical" refers to Eq.\eqref{eq:fidelity_eigenvalues} and ``Analytical with correction" to prediction taking into account localization properties (see the main text). Note the very low infidelity obtained already for this shallow circuit. \textbf{b)} The numerical infidelity of the 1-layer quantum circuit for the truncated L\'evy distribution with parameter $c=2^5$ as a function of the support interval $L$. The discretization step size is fixed, with the qubit number given by $N=10+\lceil\log_{2}L\rceil$.
       }
\end{figure*}

\section{Constructing the encoding circuit using the MPS representation}\label{sec:implementation}

We now assume an MPS encoding of a smooth function is provided, and describe how to construct a shallow (using $O(N)$ gates) encoding quantum circuit taking advantage of the above results. For small qubit numbers (less than 30) a vector of discretized function values can be converted to MPS using the standard SVD method. For large scales we employ the TCI algorithm \cite{6076873_oracle}, which is sampling based and \emph{does not} require explicit storage of the full inputs in the classical memory. This TCI version is compute- and memory-efficient allowing us to generate utility scale encoding circuits, which we executed on QPU for up to 156 qubits (see Figs.\ref{fig:big_distribution_encoded1},\ref{fig:big_distribution_encoded2}).

\begin{figure*}[t]
    \centering
    \subfigure{\includegraphics[width=0.49\textwidth]{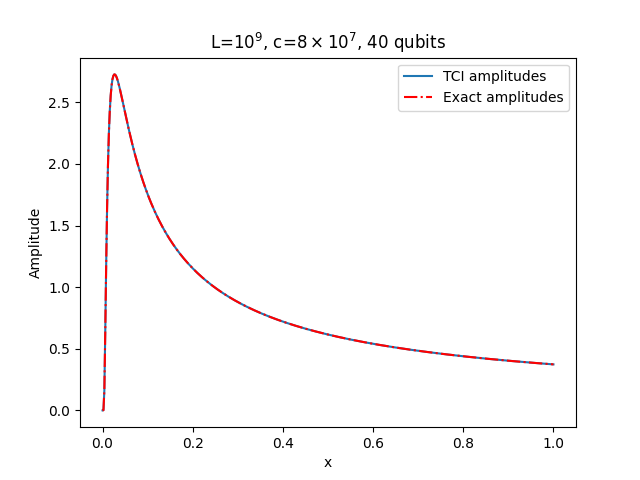}}
    \subfigure{\includegraphics[width=0.49\textwidth]{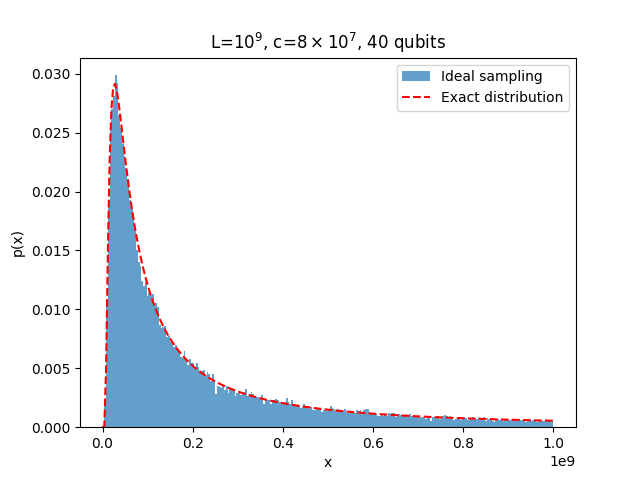}}
    \caption{\textbf{(left)} The amplitudes of the quantum state encoding the L\'evy distribution on 40 qubits obtained using Tensor Cross Interpolation (TCI), compared to the exact Matrix Product State (MPS) amplitudes \textbf{(right)} The results of sampling from the 40-qubit quantum circuit constructed using TCI approach on an ideal simulator.}
    \label{fig:tci_encoding}
\end{figure*}

Our base algorithm extends the approach of Refs.\cite{Gundlapalli_2022,Iaconis_2024, Melnikov_2023, sano_2024,GonzalezConde_2024}, which we now briefly review. Recall first that a MPS of bond dimension $\chi=2$ can be converted exactly, in a purely analytic procedure, into a quantum circuit utilizing only 2-qubit gates \cite{Ran_2020, Rudolph_2024}. Though for larger bond dimensions an exact procedure requires arbitrarily large multi-qubit gates, and is thus infeasible, an approximate procedure is constructed as follows. \textbf{(i)} The input MPS state is \emph{truncated} to bond dimension $\chi=2$. The truncation can be done by \emph{e.g.}~setting to zero all but the two largest elements of the diagonal $\Lambda$ matrices in the canonical form representation of MPS (see Eq.~\ref{eq:canonical}). \textbf{(ii)} The MPS with bond dimension $\chi=2$ can be converted exactly into a quantum circuit.  \textbf{(iii)} The hermitian conjugate of this quantum circuit is contracted with the original MPS input state (at full bond dimension) on an MPS quantum circuit simulator (we used the Qiskit MPS simulator \cite{qiskit2024}) \emph{distentangling} it. \textbf{(iv)} The procedure is iterated with the updated MPS state. Note that this approach is approximate and greedy. As MPS of bond dimension 2 is represented exactly by a quantum circuit, contracting it with its hermitian conjugate yields a product state, removing all entanglement. For MPS of higher bond dimension, contracting hermitian conjugate of the circuit obtained from a truncated representation removes the leading part of the state, and yields a state of lower entanglement, requiring a lower bond dimension for the next iteration. In practice, this method generally converges to the exact circuit for state preparation for a large number of iterations.

We improve upon the above procedure in two distinct ways. First, motivated by our theoretical results, we modify it to take advantage of the structure of the entanglement in the input MPS. Second, to further improve the practical performance we note that specific operators in the procedure admit more efficient gate decompositions than used in \cite{Gundlapalli_2022,Iaconis_2024, Melnikov_2023, sano_2024,GonzalezConde_2024}, resulting in overall shallower circuits.

\begin{figure*}[t]
    \centering\includegraphics[width=0.85\textwidth]{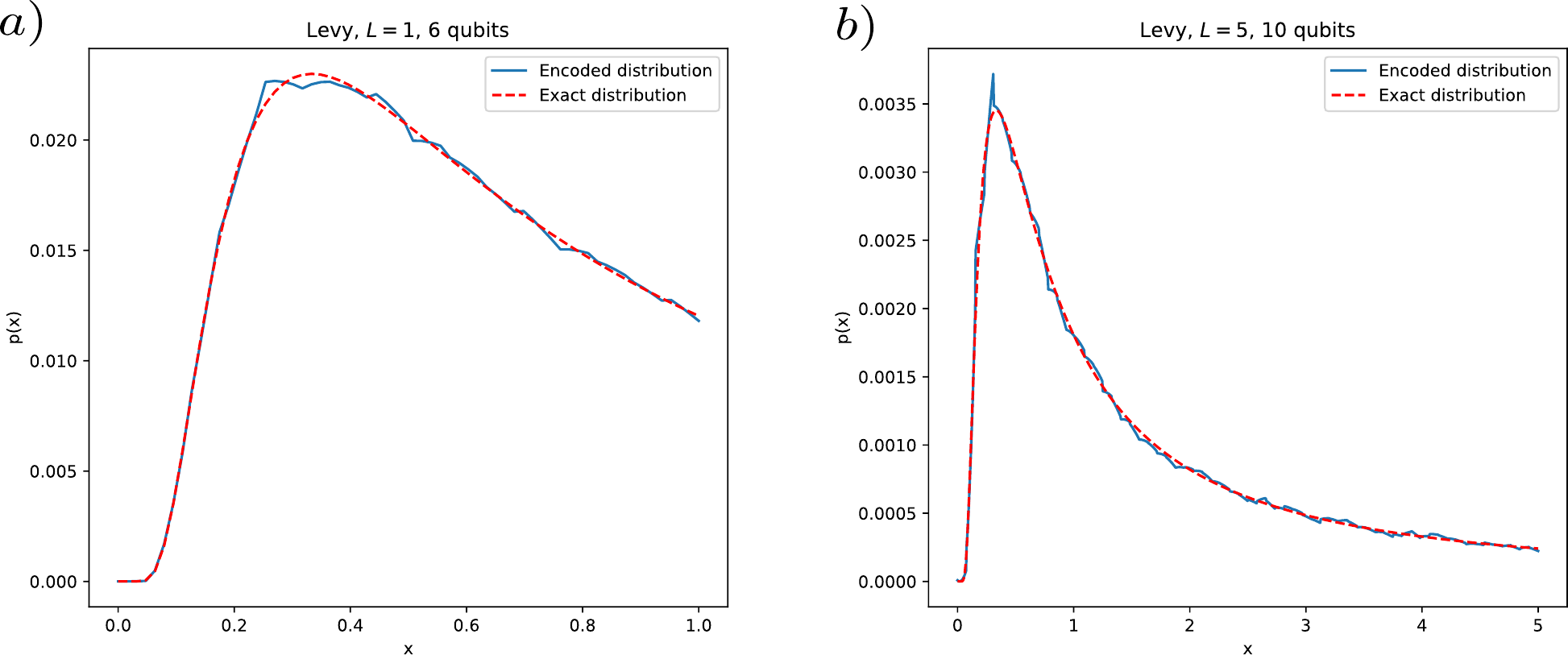}
    \caption{The encoded L\'evy distribution ($c=1$) for different truncated support and discretization (number of qubits) sizes on an ideal noiseless simulator. Amplitudes of the quantum state created using a 2-layer encoding circuit are compared to the exact distribution. In \textbf{a)} for $N=6$ qubits and $L=1$; in \textbf{b)} for $N=10$ qubits and $L=5$. }
    \label{fig:levy_distribution_encoded}
\end{figure*}

For the former, note that the exact procedure of Refs.\cite{Ran_2020, Rudolph_2024}, used also in all subsequent works, exclusively produces complete linear staircase layers as disentangling circuits. Each layer, multiple of which are generated in disentangling iterations, uses $N-1$ gates on $N$ qubits, and has a depth also of $N-1$. While this already produces a circuit with $O(N)$ gates, we show that an improved ``double-staircase" or ``V"-layer can also be generated (see Fig.\ref{fig: analytic_decomposition}), which has two advantages: first, the layer depth can be as low as $N/2 -1$ for a centrally placed origin, which is practically a very significant advantage on real noisy QPUs. Second, since, as we demonstrated, the entanglement \emph{is not} spatially uniform across the bonds, the greedy iterative disentangling procedure may in practice produce a better approximation when equipped with a freedom to move the origin of disentangling layers, even in the absence of noise. This is indeed the case, as we demonstrate numerically in Fig.\ref{fig: Levy_centre_position2}, where up to three layers are used to approximate the L\'evy distribution, and the error of approximation is plotted as a function of the layer origin. Note that the gain due to shifting the origin can be as large as adding another disentangling layer in the standard staircase scheme, thus drastically reducing the number of gates and the circuit depth.

V-layers are generated as follows: first, a central bond is chosen and the MPS is converted to the central canonical form with respect to this bond, so that the tensors to the left and to the right of the central bond are isometries (see Eq.~\ref{eq:center_canonical} and Step 0 in Fig.\ref{fig: analytic_decomposition}). The central bond can be chosen either to minimize the depth of the resulting V-layer, or it can be placed in the region with largest bond entanglement. The MPS is then truncated to bond dimension 2. For this truncation we can either briefly restore the full canonical form and set all but the two largest diagonal elements in all $\Lambda$ matrices to  zero, or we can truncate directly in the center canonical form by moving the position of central bond and sequentially truncating the $\Lambda$ matrix of the bond (the latter method is described in detail in Ref.~\cite{SCHOLLWOCK_2011}). In either case the resulting left- and right- tensors $A_{L}$ and $A_{R}$ are isometries, as defined in Eqs.~\ref{right_canonical} and ~\ref{left_canonical}, of dimension $2 \times \chi = 4$ by $\chi=2$. The isometries can be completed into unitary matrices of dimension $4 \times 4$, and interpreted as explicit matrix representations of 2-qubit gates (as in Ref.\cite{Ran_2020}). Completion here means that the resulting 2-qubit gates reduce to the original isometries if one of the qubits on which gate acts is in the $|0\rangle$ state. This is shown in in Step 2 in Fig.~\ref{fig: analytic_decomposition}, where the unitary matrix $u_{L,1}$ is obtained from the completion of the contraction of tensors $A_{L,1}$ and $A_{L,2}$, $u_{L,2}$ is found as a completion of $A_{L,3}$, $u_{R,4}$ is the completion of $A_{R,4}$ and $u_{R,5}$ is the completion of contraction of $A_{R,5}$ and $A_{R,6}$. 

It is also necessary to prepare the initial state $\sum_{i} \Lambda_{i} |ii\rangle$ on the central bond (the gate $u_{\Lambda}$ in Fig.\ref{fig: analytic_decomposition}). This requires a distinct approach. In the first layer $u_{\Lambda}$ acts on the $| 00\rangle$ state, and can be constructed using a single CNOT \cite{Iten_2016}. In subsequent layers we make this (and only this) gate variational. As for general 2-qubit isometries only up to two CNOT gates are needed \cite{Iten_2016}, we take such an ansatz with 1-qubit parametrized rotations. Their parameters are fixed by minimizing the second Renyi entropy on the central bond of the contracted disentangled MPS \cite{Hauschild_2018}.

\begin{figure*}[t]
    \centering
    \includegraphics[width=0.9\textwidth]{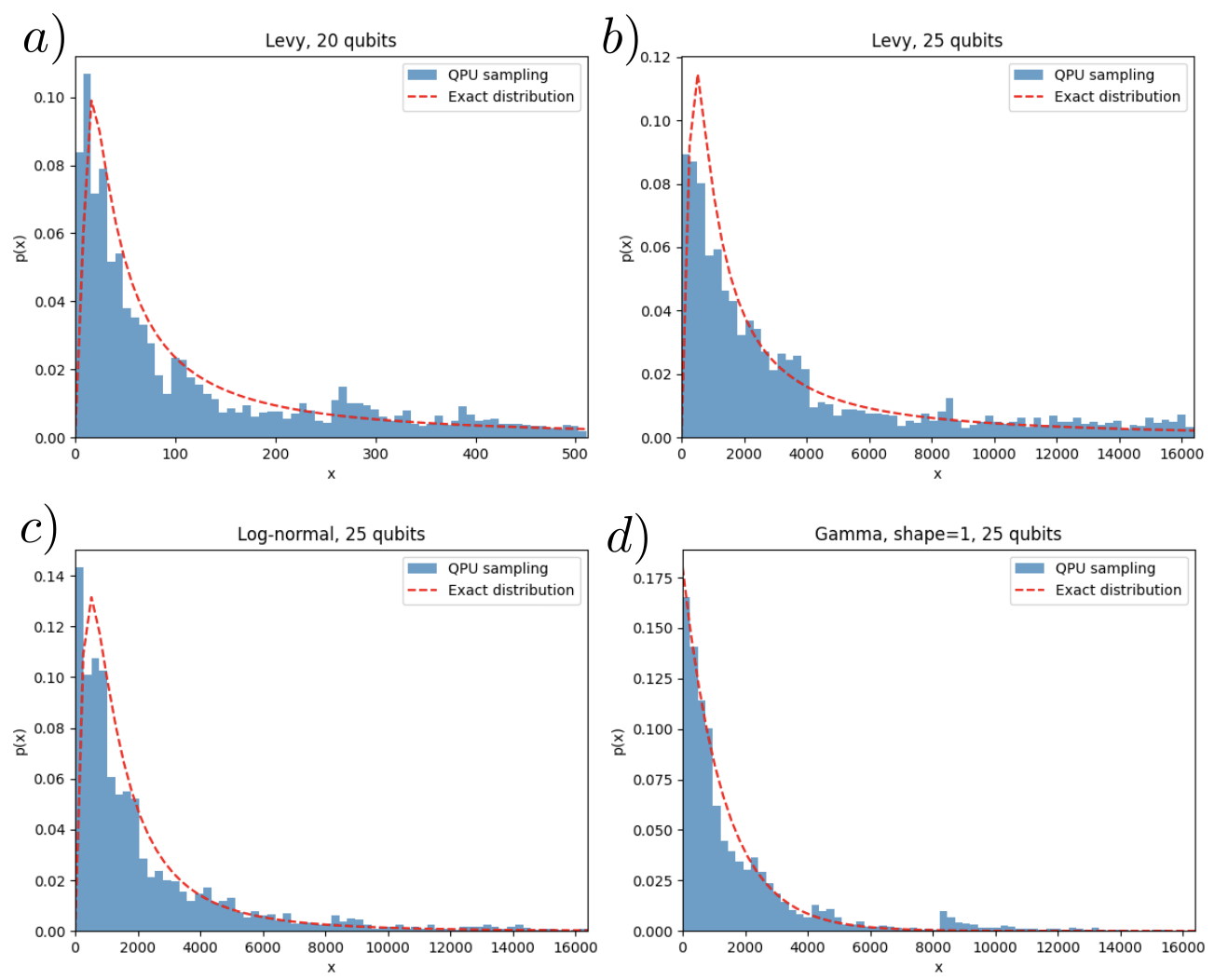}
    \caption{Encoding circuits execution on the \texttt{ibm\_torino} QPU. \textbf{a)} L\'evy distribution with $c=50$ truncated to $L = 2^9$ with $N=20$ qubits \textbf{b)} L\'evy distribution with $c=1310$ on $L=2^{14}$ with $N=25$ qubits \textbf{c)} Log-normal distribution with on $N=25$ qubits \textbf{d)} Gamma distribution with shape parameter $k=1$ on $N=25$ qubits. In each case 5000 shots were used. All executions pass the KS test with 200 samples. The better of the 1- and 2-layer circuits used is shown.}
    \label{fig:qpu_levy_large_results}
\end{figure*}

The above procedure generates ``full" disentangling V-layers with $N-1$ gates each. This is, however, not necessary as we have shown that the entanglement decays exponentially fast with bond numbers. To take advantage of this fact we introduce a hyperparameter $\varepsilon_{trunc}>0$, on whose value we condition the addition of gates on individual bonds in the V-layer. Namely, for each bond $k$ we compute $\sum_{i\geq 1} \Lambda_{k,i}^{2}$ in the contracted MPS. If this is smaller then the chosen small $\varepsilon_{trunc}$ then the state is essentially already disentangled, and the bond can be truncated to $\chi=1$, \emph{i.e.}~no new gate is needed. In this way in subsequent iterations we generate ``incomplete" and diminishing V-layers, as depicted in Fig.\ref{fig:mainfig}d, resulting in overall fewer gates.

We have also modified  the gate synthesis procedure for the isometric tensors. In the standard procedure the 2-qubit isometry is extended into a full 2-qubit unitary matrix, from which the gate representation is synthesized. Such a general unitary synthesis requires, however, 3 CNOT gates and is not optimal. In fact, to synthesize a general isometry only 2 CNOTs are needed and 1-qubit RY and RZ gates \cite{Iten_2016}, which we use, further reducing the overall and 2-qubit gate count. For real-valued functions one may even further simplify the circuits by only using the RY rotations \cite{Wei_2012}).

\section{Numerical tests of the encoding circuits on simulators and QPUs}\label{sec:experiments}

Here we numerically study the quantum circuits constructed with our MPS based procedure both on ideal and noisy simulators, as well as on real IBM QPUs.

To measure the accuracy of the prepared quantum circuit we use the fidelity defined as $F = |\langle \psi| f \rangle|^{2} $, where $|f\rangle$ is the exact state and $|\psi\rangle$ is the state prepared by the circuit. For a single-layer disentangling circuit we can analytically predict the fidelity of the approximation using the truncated singular values. Truncating a single MPS bond $k$ to $\chi=2$ results in fidelity $1 - \sum_{i\geq 2} \Lambda_{k,i}^{2}$, where $\Lambda_{k,i}$ are the entanglement spectra/singular values on the $k$-th bond. With multiple bonds truncated, to lowest order the fidelity is:
\begin{equation}\label{eq:fidelity_eigenvalues}
    F \approx 1 - \sum_{\substack{k>0, i\geq2}} \Lambda^{2}_{k,i},
\end{equation}
which can be effectively computed up to an exact leading term via Theorem~\ref{thm:density_matrix} (see Appendix~\ref{app:B} for details). The multilayer decomposition does not easily admit such analysis, and we only show numerical results.

In Fig.~\ref{fig: normal_error}a we show the infidelity of a 1-layer approximating quantum circuit as a function of the truncated support size $L$, assuming a fixed discretization size (thus the qubit number is $N=10+\lceil\log_{2}L\rceil$). Here we consider the example of a normal distribution with $\sigma=2$. The infidelity is computed on an MPS simulator with a very large bond dimension (\emph{i.e.}~essentially exact simulation). Overall, the infidelity is excellent already for this extremely shallow circuit, no higher than $10^{-3}$. Note also that for smaller $L$ it perfectly follows the analytical estimate from Eq.\eqref{eq:fidelity_eigenvalues}. For larger supports this simple approximation is not accurate: as  discussed in Sec.\ref{sec:generalres} the smoothness-controlled regime is valid from bond $k = \log_{2}(L/2\sigma)$ upwards, and below the entanglement is localization controlled (and, for this exponentially localized function, much smaller, see Fig.\ref{fig2: localization}a). For large $L$ thus Eq.\eqref{eq:fidelity_eigenvalues} overestimates the infidelity, which can be corrected by only summing contributions from bonds $k \geq log_{2}(L/2\sigma)$, \emph{i.e.}~the ones for which the asymptotic scaling is valid. This corrected estimate agrees perfectly with the numerical results for all $L$.

\begin{table*}[t]
\begin{tabular}{c|c|c|c|c|c|}
\cline{2-6}
                                                  & Qubits & \begin{tabular}[c]{@{}c@{}}KL divergence\\ \textit{mean$\pm$std}\end{tabular} & \begin{tabular}[c]{@{}c@{}}KS test p-value\\ \textit{mean$\pm$std}\end{tabular} & \begin{tabular}[c]{@{}c@{}}Depth\\ \textit{mean}\end{tabular} & \begin{tabular}[c]{@{}c@{}}CNOTs\\ \textit{mean}\end{tabular} \\ \hline
\multicolumn{1}{|c|}{\multirow{3}{*}{Normal}}     & 6      & 0.00031 $\pm$ 0.00048                                                & 0.23839 $\pm$ 0.23489                                                  & 17.2                                                 & 9.8                                                  \\
\multicolumn{1}{|c|}{}                            & 10     & 0.00045 $\pm$ 0.00069                                                & 0.49284 $\pm$ 0.28943                                                  & 26.2                                                 & 17.8                                                 \\
\multicolumn{1}{|c|}{}                            & 15     & 0.00046  $\pm$ 0.00073                                               & 0.50486 $\pm$ 0.28203                                                  & 41.2                                                 & 28.0                                                 \\ \hline
\multicolumn{1}{|c|}{\multirow{3}{*}{Log-normal}} & 6      & 0.00097 $\pm$ 0.00075                                                & 0.08986 $\pm$ 0.16966                                                  & 15.6                                                 & 9.0                                                  \\
\multicolumn{1}{|c|}{}                            & 10     & 0.00148 $\pm$ 0.00135                                                & 0.31905 $\pm$ 0.32349                                                  & 28.4                                                 & 18.2                                                 \\
\multicolumn{1}{|c|}{}                            & 15     & 0.00115 $\pm$ 0.00112                                                & 0.47337 $\pm$ 0.28552                                                  & 40.2                                                 & 27.0                                                 \\ \hline
\multicolumn{1}{|c|}{\multirow{3}{*}{L\'evy}}       & 6      & 0.00175 $\pm$ 0.00086                                                & 0.26624 $\pm$ 0.22399                                                  & 15.6                                                 & 9.0                                                  \\
\multicolumn{1}{|c|}{}                            & 10     & 0.00172 $\pm$ 0.00053                                                & 0.50047 $\pm$ 0.27975                                                  & 25.9                                                 & 17.0                                                 \\
\multicolumn{1}{|c|}{}                            & 15     & 0.00149 $\pm$ 0.00048                                                & 0.50345 $\pm$ 0.28411                                                  & 41.0                                                 & 27.0                                                 \\ \hline
\end{tabular}
\caption{Statistical tests of the encoded distribution on an ideal simulator and the encoding circuit's (two-layers) properties with a $\varepsilon_{trunc}=10^{-3}$ cutoff.}
\label{ta:eps0.001}

\bigskip
\begin{tabular}{c|c|c|c|c|c|}
\cline{2-6}
                                                  & Qubits & \begin{tabular}[c]{@{}c@{}}KL divergence\\ \textit{mean$\pm$std}\end{tabular} & \begin{tabular}[c]{@{}c@{}}KS test p-value\\ \textit{mean$\pm$std}\end{tabular} & \begin{tabular}[c]{@{}c@{}}Depth\\ \textit{mean}\end{tabular} & \begin{tabular}[c]{@{}c@{}}CNOTs\\ \textit{mean}\end{tabular} \\ \hline
\multicolumn{1}{|c|}{\multirow{3}{*}{Normal}}     & 6      & 0.00023 $\pm$ 0.00039                                                & 0.24417 $\pm$ 0.24345                                                  & 18.2                                                 & 10.4                                                 \\
\multicolumn{1}{|c|}{}                            & 10     & 0.00038 $\pm$ 0.00063                                                & 0.49001 $\pm$ 0.28571                                                  & 27.5                                                 & 18.7                                                 \\
\multicolumn{1}{|c|}{}                            & 15     & 0.00037 $\pm$ 0.00061                                                & 0.50592 $\pm$ 0.28207                                                  & 41.6                                                 & 29.2                                                 \\ \hline
\multicolumn{1}{|c|}{\multirow{3}{*}{Log-normal}} & 6      & 0.00076 $\pm$ 0.00064                                                & 0.09607 $\pm$ 0.17782                                                  & 19.5                                                 & 12.1                                                 \\
\multicolumn{1}{|c|}{}                            & 10     & 0.00127 $\pm$ 0.00120                                                & 0.32650 $\pm$ 0.32761                                                  & 31.1                                                 & 21.8                                                 \\
\multicolumn{1}{|c|}{}                            & 15     & 0.00106 $\pm$ 0.00108                                                & 0.47656 $\pm$ 0.29033                                                  & 43.8                                                 & 32.7                                                 \\ \hline
\multicolumn{1}{|c|}{\multirow{3}{*}{L\'evy}}       & 6      & 0.00101 $\pm$ 0.00097                                                & 0.26329 $\pm$ 0.22342                                                  & 20.6                                                 & 11.8                                                 \\
\multicolumn{1}{|c|}{}                            & 10     & 0.00112 $\pm$ 0.00073                                                & 0.49681 $\pm$ 0.27833                                                  & 30.5                                                 & 20.8                                                 \\
\multicolumn{1}{|c|}{}                            & 15     & 0.00099 $\pm$ 0.00064                                                & 0.49213 $\pm$ 0.28745                                                  & 42.8                                                 & 31.8                                                 \\ \hline
\end{tabular}
\caption{Statistical tests of the encoded distribution on an ideal simulator and the encoding circuit's (two-layers) properties with a $\varepsilon_{trunc}=10^{-4}$ cutoff.}
\label{ta:eps0.0001}
\end{table*}

In Fig.~\ref{fig: normal_error}b we show the numerical infidelity of the 1-layer quantum circuit for the L\'evy distribution as a function of the truncated support $L$. We used a fixed discretization step ($N=10+\lceil\log_{2}L\rceil$ qubits). Note again the very low infidelity achieved already with a single layer of gates, saturating at $4 \times 10^{-3}$. While for this case the analytical estimates based on Eq.\eqref{eq:fidelity_eigenvalues} are less reliable, L\'evy distribution being only weakly localized, the qualitative behaviour of infidelity saturation at large $L$ is expected: The non-universal contributions from a (finite) set of larger scales giving place to smoothness controlled regime at larger scales where the entanglement systematically vanishes.

In Fig.\ref{fig: Levy_centre_position2} we examine the influence of using multiple disentangling V-layers in the quantum circuit construction, and the sensitivity to the V-layer origin position, on the example of the L\'evy distribution. We observe that the fidelity can be systematically improved by adding additional disentangling layers (note, though, that in the presence of noise this is not true, as the increased circuit depth may result in a lower fidelity in practice). Furthermore, for more than a single layer we also see a strong dependence on the layer origin, as expected from the entanglement analysis in Sec.\ref{sec:generalres}: the effect is comparable in size to adding another disentangling layer, and hence the freedom to chose it is very important in practice. Note also that the minimal circuit depth is achieved with the origin on the middle qubits, further emphasizing the usefulness of this strategy.

We show the encoded distributions themselves in Fig.\ref{fig:levy_distribution_encoded}, where the amplitudes of the state created with a 2-layer disentangling circuit on an ideal simulator are compared to the exact distributions for different support intervals $L$ and discretizations (specified by the number of qubits $N$). We observe a very good agreement, and emphasize that this is achieved using only linearly deep circuits, which is crucial when considering execution on realistic noisy devices.

\begin{figure*}[t]
    \centering
    \includegraphics[width=\textwidth]{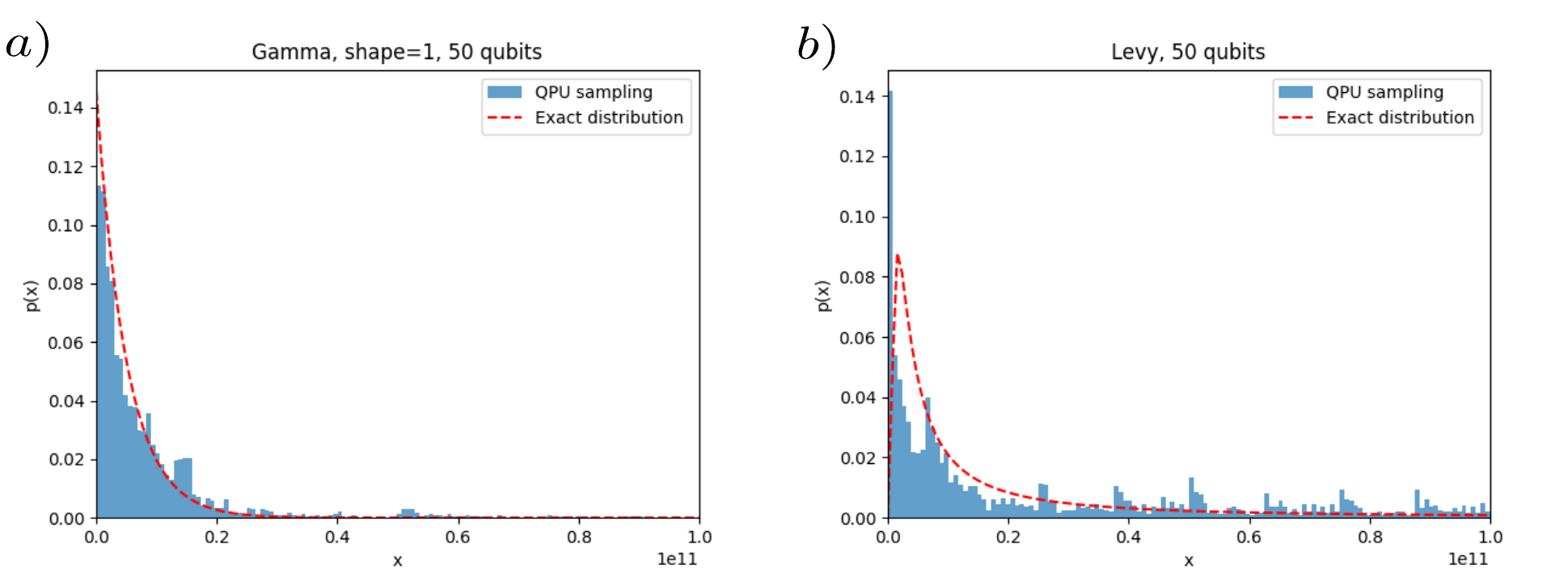}
    \caption{Tensor Cross Interpolation (TCI) constructed 2-layer encoding circuits for the \textbf{a)} Gamma and \textbf{b)} L\'evy distributions on 50 qubits, executed on the \texttt{ibm\_marrakesh} QPU. We used in both cases $L=10^{11}$, the scale factor $c=5\times10^9$, and 5000 shots on the QPU. While these very large circuits no longer pass the quantitative statistical tests, the qualitative features of the distributions are still reproduced. Compare, however, the results on higher quality QPU in Fig.\ref{fig:big_distribution_encoded2}.}
    \label{fig:big_distribution_encoded1}
\end{figure*}

\begin{figure*}[t]
    \centering
    \subfigure{\includegraphics[width=0.325\textwidth]{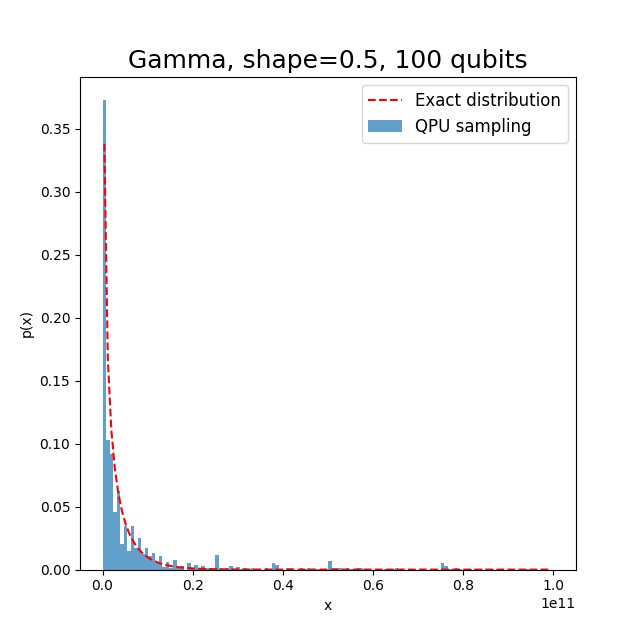}}
    \subfigure{\includegraphics[width=0.325\textwidth]{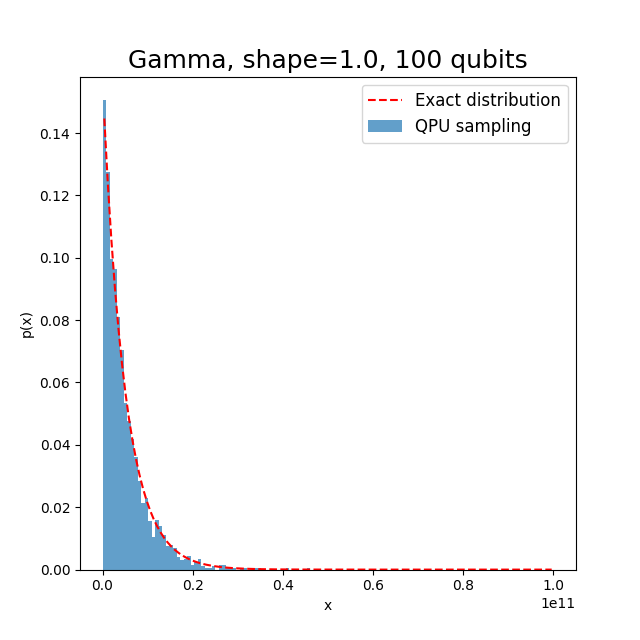}}
    \subfigure{\includegraphics[width=0.325\textwidth]{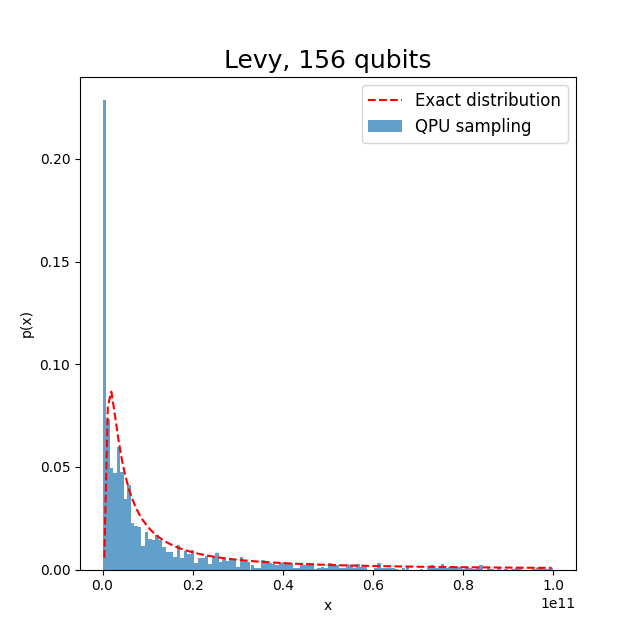}}
    \caption{Sampling from 2-layer encoding circuits on the \texttt{ibm\_kingston} QPU. We used Tensor Cross Interpolation (TCI) in circuit generation, and measured 2000 shots on the QPU; we also used dynamical decoupling. In all plots we used $L=10^{11}$, and the scale factor $c=5\times10^9$. \textbf{(left)} Gamma distribution with shape parameter $a=1$ sampled on 100 qubits. \textbf{(middle)} Gamma distribution with shape parameter $a=0.5$ sampled on 100 qubits. \textbf{(right)} L\'evy distribution sampled on 156 qubits. Note that while both \texttt{ibm\_marrakesh} and \texttt{ibm\_kingston} are QPUs of type Heron r2, the latter has a significantly lower 2-qubit gate error (2.67e-3 \emph{vs.}~1.88 e-3 median error), which is reflected in the much higher quality of results compared to Fig.\ref{fig:big_distribution_encoded1}.}
    \label{fig:big_distribution_encoded2}
\end{figure*}

A more rigorous test of performance is obtained by computing statistical metrics for the samples drawn from the encoded distributions. We use the Kullback-Leibler (KL) divergence between the encoded and ideal distributions, and the Kolmogorov-Smirnov (KS) test. The results are reported in Tables \ref{ta:eps0.001} and \ref{ta:eps0.0001}. For these tests we use a 2-layer encoding circuit constructed by our complete algorithm as described in Sec.\ref{sec:implementation}, including truncation of gates in V-layers set by the parameter $\varepsilon_{trunc}$. We report the depth and amount of CNOT gates of the circuit, which are computed in the RX, RY, RZ and CNOT basis (the depth includes all one-qubit gates). To collect the data we varied the interval $L$ and scale factors of the distributions, and we report the mean value and the standard deviation. The constructed distributions mostly pass the statistical tests with a wide margin, but we note that for very small amount of qubits KS test may occasionally fail (p-value$<0.05$) due to extreme discretization of the distribution, as the test is performed against the ideal truncated cumulative density function which is continuous. In Table~\ref{ta:eps0.001} we set the hyperparameter $\varepsilon_{trunc}$ to $10^{-3}$, while in Table~\ref{ta:eps0.0001} to $\varepsilon_{trunc}$ to $10^{-4}$. Changing this value represents an accuracy-depth trade-off, which will be important on noisy QPUs.

To emphasize the above mentioned depth-accuracy trade-off induced by realistic noise we perform noisy simulations on the IBM \texttt{FakeToronto} backend. We compare the results of the execution of a 1-layer and 5-layer encoding circuits in Fig.~\ref{fig:Levy_scaled_noise}. While in the absence of noise the latter is more accurate, in this more realistic setting the distribution features are largely destroyed by the noise, while the nominally less accurate 1-layer encoding performs much better. We leave detailed examination of this trade-off to future work.

Finally, we put our encoding circuits to test on real QPUs. Motivated by the observed theoretical accuracy-depth trade-off we executed  shallow 1- and 2- layer circuits generated by our procedure ($\varepsilon_{trunc}=10^{-4}$) on the \texttt{ibm\_torino} superconducting device. In Fig.~\ref{fig:qpu_levy_large_results} 
we show the distributions sampled from the better of the two encoding circuits for the L\'evy, Log-normal and Gamma distributions for up to $N=25$ qubits. For each plot 5000 shots we used. All these experiments pass the KS test at the standard significance level $0.05$ (with 200 samples).

Furthermore, using the TCI approach, we also generate large scale circuits, for up to 156 qubits, which we execute on IBM devices. First, we validate the TCI function loading itself. Towards this goal we generated a 40-qubit encoding circuit for the L\'evy distribution and interval size $L=10^9$, using the TCI algorithm of Ref.\cite{6076873_oracle} and \texttt{torchTT} package \cite{TorchTT} to generate the MPS representation directly from the PDF. Specifically, we build a map $[0,1]\mapsto \sqrt{L\cdot p_{Levy}(xL)} / R$, where $\sqrt{L}$ accounts for the renormalization after changing the variables, and a constant factor $R$ accounts for a truncation of the distribution to a finite interval and can be factored out of the CDF. In Fig.~\ref{fig:tci_encoding} we show the encoded state, and sampling results from a 40-qubit encoding quantum circuit executed on an ideal simulator. The large scaling factor $c$ is chosen so that the distribution tail does not vanish in the truncated range. Numerically we confirm that the relative error between the true PDF and the TCI uploaded value in the MPS is $6\times 10^{-8}$ on average, and $4\times 10^{-7}$ at most. We note that $8\times10^4$ calls to the PDF map sufficed to built the state to this accuracy.

Next, in Fig.\ref{fig:big_distribution_encoded1} we show the results of executing the TCI generated 2-layer quantum circuits for the Gamma and L\'evy disributions at 50 qubits, executed on the \texttt{ibm\_marrakesh} QPU with 5000 shots. While these large scale distributions no longer pass the statistical tests due to device noise, the qualitative features of the distributions are reproduced. To showcase how the improvement in device quality directly impacts the sampling results, in Fig.\ref{fig:big_distribution_encoded2} we show the results on the \texttt{ibm\_kingston} QPU. While this is also a Heron r2 type device, its 2-qubit gate fidelities were substantially better (1.88e-3 \emph{vs.}~2.67e-3 for \texttt{ibm\_marrakesh}). This allowed to sample from Gamma and L\'evy distribution on up to 156 qubits, which we performed using 2000 measurement shots.

\section{Conclusions}\label{sec:conclusions}

We studied entanglement properties of MPS representations of smooth functions, and their applications to constructing shallow encoding quantum circuits. In particular, we derived the universal properties of their entanglement spectra. We used these to estimate the errors of the quantum circuit representations of these functions, and to build an improved MPS-based algorithm generating shallow quantum circuits [$O(N)$ gates] encoding them. The resulting circuits were benchmarked on real and simulated IBM QPUs. The practical performance of our algorithm was validated experimentally on the important example of preparing states encoding probability distribution functions (PDF), in particular heavy-tailed ones. Both the analytical and algorithmic results apply, however, to general smooth functions. One further domain where this may be of importance is the preparation of initial states for quantum simulation of partial differential equations (PDE), used \emph{e.g.}~in quantum computational fluid dynamics (CFD) \cite{ljubomir2022quantum}.

For the important case of L\'evy distributions the quantum circuits we constructed reproduced well their qualitative behaviour for up to 156 qubits. Moreover, for up to 25 qubits the accuracy was sufficient to pass quantitative statistical tests even on the current noisy devices. The results on distribution loading may have importance to various domains of financial risk analysis, risk management, and decision-making that includes series of financial data. By developing methods to efficiently work with L\'{e}vy distributions on a quantum computer, we pave the way for more precise modeling of market behaviors, particularly in capturing heavy tails, skewness, and volatility clustering.

It is of key importance that our procedure does not require the explicit storage of the input, \emph{i.e.}~the potentially exponentially many discretized function values, in memory. Instead, it is able construct the circuit directly, by locally sampling from the probability density function (PDF) trough the use of Tensor Cross Interpolation (TCI)\cite{TCI_paper}. Thus our method is both classical compute- and memory-efficient, and scalable to utility sizes.

Several open research directions remain. First, the density matrix based analysis can be extended to multivariate functions. It is particularly interesting to derive the optimal tensor network representation of such functions based on their entanglement (for multivariate functions there are many possible tensor network geometries, which may represent the function; from the point of quantum computing these geometries correspond to different qubit orderings, or qubits arranged in topologies more complicated than linear). Analytic results on entanglement of multivariate functions may lead both to better quantum circuit encodings of such functions, and also may improve existing tensor network algorithms. Finally, from a practical stand point it would also be very important to carry out a more detailed analysis of the influence of hardware noise on the function preparation. 

\section*{Acknowledgements}
We thank Vladyslav Los and Oleksandra Hutor for their technical support in executing the quantum circuits on the IBM quantum hardware.
\\
\\
\textbf{Disclaimer.} This paper was prepared for information purposes and is not a product of HSBC
Bank Plc.~or its affiliates. Neither HSBC Bank Plc.~nor any of its affiliates make any explicit
or implied representation or warranty and none of them accept any liability in connection with
this paper, including, but not limited to, the completeness, accuracy, reliability of information
contained herein and the potential legal, compliance, tax or accounting effects thereof. Copyright
HSBC Group 2024




\newpage
\appendix
\section{Proofs of theorems}
\label{app:A}

\subsection{Proof of Theorem \ref{thm:purities}}

To obtain the purities we first construct the reduced density matrix $\rho_{k}$ in terms of the tensor representation:
\begin{equation}
    \rho_{k} = \frac{1}{2^{N}} \sum_{\mathbf{v_{k}}} f_{\mathbf{u_{k}}, \mathbf{v_{k}}} f^{*}_{\mathbf{u_{k}'}, \mathbf{v_{k}}},
\end{equation}
where we introduced collective indices $\mathbf{u_{k}} = \sigma_{1} \sigma_{2}\dots\sigma_{k}$, and $\mathbf{v_{k}} = \sigma_{k+1} \sigma_{k+2}\dots\sigma_{N}$ for the bipartition of the system (across spatial scales). 
Here $f^*$ denotes a complex conjugate of $f$.
Then the purity is given by:
\begin{equation}
    p_{k} = \frac{1}{4^{N}} \sum_{\mathbf{u_{k}, u'_{k}, v_{k}, v'_{k}}} f_{\mathbf{u_{k},v_k}} f^{*}_{\mathbf{u_k, v'_k}} f^{*}_{\mathbf{u'_{k}, v_{k}}} f_{\mathbf{u'_{k}, v'_{k}}}.
\end{equation}

We further write the value $f_{\mathbf{u_{k}, v_{k}}}$ as $f(\mathbf{u_{k}} + \mathbf{v_{k}}/2^{k})$, where, by a slight abuse of the notation, in the argument of the function f we have identified the index $\mathbf{u_k}$ with the real number $0.\mathbf{u_k} \equiv 0.\sigma_1\ldots\sigma_k$, and $\mathbf{v_k}$ with $0.\mathbf{v_k} \equiv 0.\sigma_{k+1}\ldots\sigma_N$. In the limit $N \to \infty$, the sums over $\mathbf{v_{k}, v'_{k}}$ may be rewritten as integrals over continuous $v, v'$:

\begin{align}
    p_{k} = \frac{1}{4^{k}} \sum_{\mathbf{u_{k}, u'_{k}}} &\int_{0}^{1} \int_{0}^{1} \left[ f(\mathbf{u_{k}} + v/2^{k}) f^{*}(\mathbf{u_{k}}+v'/2^{k}) \right. \nonumber \\ &\left. f(\mathbf{u'_{k}}+v'/2^{k}) f^{*}(\mathbf{u'_{k}} + v/2^{k}) \right] \rm{d} v \rm{d}v'.
\end{align}

The sums over $\mathbf{u_{k}, u'_{k}}$ cannot be equally converted to integrals by taking the limit $k \to \infty$, as we want the purities at finite $k$. To this end we use the Euler-MacLaurin resummation formulas, which approximate the finite $1/2^{k}$ resolution sum with continuous integrals augmented with additional boundary terms. In our specific case we use:

\begin{align}
    \sum_{u_{n}} \frac{f(u_{n})}{2^{k}} &= \int_{0}^{1} f(u) \rm{d} u + \\ \nonumber &+ \frac{f(0) - f(1)}{2^{k+1}} + B_{2} \frac{f'(1) - f'(0)}{2^{2k+1}} + O(1/2^{3k}),
\end{align}
where $B_{2}$ is the second Bernoulli number. Then the purity is given by:
\begin{multline}
    p_{k} = \int_{0}^{1} \int_{0}^{1} \left(\int_{0}^{1} f(u + \frac{v}{2^{k}}) f^{*}(u+ \frac{v'}{2^{k}}) \rm{d} u + b_{1}(v, v') \right) \\ \left(\int_{0}^{1} f^{*}(u' + \frac{v}{2^{k}}) f(u'+ \frac{v'}{2^{k}}) \rm{d} u' + b_{2}(v, v') \right) \rm{d} v \rm{d} v',
    \label{eq:purity}
\end{multline}
where $b_{1}, b_{2}$ are the boundary terms. There is a small subtlety here, as the function here is effectively extended beyond $[0,1]$, to the interval $[0, 1+1/2^{k}]$. We assume that such an extension exists and is smooth. Indeed any smooth function on $[0,1]$ can be easily extended. More over, on practice it's likely that the function is already defined beyond this interval which is used for encoding.

To evaluate the $v, v'$ integrals, for large $k$ we can expand in the small parameter $v/2^{k}$: 
\begin{equation}f(u + v/2^{k}) = f(u) + f'(u) \frac{v}{2^{k}} + f''(u) \frac{v^{2}}{ 2^{2k+1}} + O(1/2^{3k})
\label{eq:fexpan}
\end{equation}
This effectively leads to an expansion in $1/2^{k}$ for the purities $p_{k}$, which we will now show.
First, the leading term in $p_k$ is equal to one, where we use the fact that boundary terms come at order at least $1/2^{k}$ and the function is normalised:
\begin{equation}
    \left(\int_{0}^{1} f(u) f^{*}(u) \rm{d}u\right) \\ \left(\int_{0}^{1} f^{*}(u') f(u') \rm{d}u' \right) = 1,
\end{equation}

\noindent
The contributions to the $p_k$ expansion at order $1/2^{k}$ come from the corresponding terms in the expansion Eq.\ref{eq:fexpan} and the boundary terms. Introducing the notation:
\begin{equation}
    h_{n,m} = \int_{0}^{1} f^{(n)}(u) f^{*(m)}(u) \rm{d}u.
\end{equation}
we now can see that the $1/2^{k}$ terms in the $p_k$ expansion vanish:
\begin{align}
    h_{0,1} + h_{1,0} + |f(0)|^{2} - |f(1)|^{2}=0,
\end{align}
since $h_{0,1} + h_{1,0}$ is the integral of a total derivative from the integration by parts formula, which cancels the boundary terms. 

Finally, we evaluate the contributions to $p_k$ at $1/4^{k}$, which come from gathering the terms in the expansion of the $du$ and $du'$ integrals in Eq.\ref{eq:purity} to second order. From a product of the 0-th order term in $u$ and second order terms in $u'$ we obtain:
\begin{multline}
    (h_{2,0} + h_{0,2})/6 + h_{1,1}/4 - (|f(1)^{2}|' - |f^{2}(0)|')/6 = \\ = (h_{2,0} + 2h_{1,1}+h_{0,2})/6 -h_{1,1}/12 - (|f(1)^{2}|' - |f^{2}(0)|')/6 \\ = - h_{1,1}/12
\end{multline}
Above, in the second line, we used the fact that $h_{2,0} + 2h_{1,1}+h_{0,2}$ is a total derivative, which cancels the boundary terms. We obtain an exactly equal contribution from the term with the roles of $u$ and $u'$ exchanged. Lastly, the contribution of a product of first order terms in both $u$ and $u'$ gives:
\begin{multline}
    2 h_{1,0} \times h_{0,1}/3 + h_{1,0}^{2} /4 + h_{0,1}^{2}/4 - (|f(0)|^{2} - |f(1)|^{2})^{2}/4 = \\ = (h_{1,0} + h_{0,1})^{2}/4 + h_{1,0}\times h_{0,1}/6 - (|f(0)|^{2} - |f(1)|^{2})^{2}/4 \\ = h_{1,0}  h_{0,1}/6,
\end{multline}
where in the second line we collected a square of the total derivative, which cancels the boundary terms. 

Gathering all of the contributions we obtain the final result for the purity $p_{k}$:

\begin{equation}
    p_{k} = 1 - \frac{g_1(f)}{6 \times 4^{k}} + O(1/8^{k}),
    \label{eq:pk1}
\end{equation}
with:
\begin{equation}
    g_1(f)\equiv  h_{1,1} - h_{1,0} h_{0,1} = \langle f^{(1)}|\hat{P}_{0}| f^{(1)} \rangle,
    \label{eq:pk2}
\end{equation}
where $\hat{P}_{0}$ is the operator, which projects out the function $f(u)$ and scalar product $\langle f^{(1)}| f^{(1)} \rangle$ should be considered to be defined consistently with the $L_{2}$ norm on the interval of function definition.
\qed

Note that the functional $g_{1}(f)$ obeys several properties, which serve as a consistency check for the $p_k$ computation. First, it is positive definite for all functions $f(x)$, which corresponds to the purities $p_{k}$ being smaller or equal to unity, as required. Second, the value is real. Finally, we may ask when it vanishes, \emph{i.e.~}when $g_{1}(f) = 0$. For this to hold, the derivative of $f(x)$ should be in the kernel of $\hat{P}_{0}$ , which implies that the derivative is proportional to the function itself.

The only solution of such differential equation is the exponential function: $f(x) = c\exp (b x)$. This is consistent with the fact that exponential functions can be exactly expressed as MPS with bond dimension equal to one \cite{khoromskij_2011}, and hence their purities must vanish because of trivial factorizability on every bond. 

Note also that reduced density matrices may be used to construct other entanglement measures. Some of them, like higher-order generalizations of purity $\mathrm{Tr}{\rho^{4}}, \mathrm{Tr}{\rho^{6}}$ are amenable to the calculations along the lines developed above.

\subsection{Proof of Theorem~\ref{thm:density_matrix}}

Let us consider the reduced density matrix of the first $k$ qubits $\rho_k$ and compute its particular element:
\begin{equation}\label{eq:eq33}
    \rho_{k}(\mathbf{u_{k}, u'_{k}}) = \frac{1}{2^{k}} \int_{0}^{1} f(\mathbf{u_{k}} + v/2^{k}) f^{*}(\mathbf{u'_{k}}+v/2^{k}) \rm{d} v,
\end{equation}
where $\mathbf{u_{k}, u'_{k}}$ are the combined indices of the reduced density matrix. Expanding the functions $f,f^*$ in $v/2^{k}$ we can write the reduced density matrix as

\begin{equation}\label{eq:density_matrix}
    \rho_{k} = \frac{1}{2^{k}}\sum_{m,n \geq 0}  \frac{|f^{(m)}(u) \rangle \langle f^{(n)}(u')|}{(1+m+n) n! m! 2^{k(m+n)}} ,
\end{equation}
here $f^{(n)}(u)$ are derivatives of order $n$, and we used the braket notation. This is a derivative expansion, but it is crucial that this is also an expansion in $1/2^{k}$, which implies that only derivatives of small order will contribute. Note that as in the previous theorem, to obtain asymptotic expansions we consider a situation where $N\to\infty$.

Note that the states $|f^{(n)}(u)\rangle$ are not normalized and not orthogonal, but can be made so using the Gram-Schmidt method. 
To this end we define projectors $P_{m}$, projecting out all derivatives up to order $m$ at consecutive steps of the Gram-Schmidt procedure.
Particularly, $P_{0}$ projects out the function itself. The new orthogonal basis is now given by: $\{|f(u)\rangle, P_{0} |f^{(1)}(u)\rangle, P_{1} |f^{(2)}(u)\rangle,\dots\}$. The basis change in the reduced density matrix can now be performed by iteratively replacing
\begin{equation}
|f^{(1)}(u) \rangle \to P_{0} |f^{(1)}(u) \rangle + (1 - P_{0}) |f^{(1)}(u) \rangle
\end{equation} 
in Eq.\ref{eq:density_matrix}, 
collecting the terms, and repeating recursively for higher derivatives. 
Note that this transformation is \emph{triangular} in a sense that after applying it only the density matrix coefficients with lower indices are affected (\emph{e.g.~}$(1 - P_{0}) |f^{(1)}(u) \rangle \propto |f(u) \rangle$ will generate correction to the $|f^{(0)}(u)\rangle$). Since the initial coefficients of $|f^{(m)}(u) \rangle  \langle f^{(n)}(u')|$ are of order $1/2^{k(m+n)}$, the orthogonalization only changes them at subleading orders in $1/2^{k}$, and can be ignored.

The new basis vectors must be normalized, which gives additional $\sqrt{\langle f^{n}(u) | P_{n-1}| f^{n}(u) \rangle}$ factors in the density matrix coefficients. After a simple but tedious calculation we obtain the following expression for $\rho_{k,m,n}$ in the orthonormal basis, to leading order in $1/2^{k}$: 
\begin{align}
& \rho_{k, m,n} = \\ \nonumber &= \frac{\sqrt{\langle f^{(n)}(u) | P_{n-1}| f^{(n)}(u) \rangle \langle f^{(m)}(u) | P_{m-1}| f^{(m)}(u) \rangle}}{(m+n+1) m! n! 2^{k(m+n)}},
\end{align}
where $m,n \in \mathbb{N}_0$ and $\rho_{k, m,n}$ is a recomputed term under the sum in Eq.\ref{eq:density_matrix}. 
The scalar products $\langle f^{(m)}(u) | P_{m-1}| f^{(m)}(u) \rangle$ appearing in the density matrix
are defined for the function discretized at a finite resolution $1/2^{k}$. However, as before, via the Euler-MacLaurin formula they can be approximated by continuous integrals with boundary terms (which, as we work to leading order in $1/2^{k}$ throughout, we ignore). In particular, we have $\langle f^{(1)}(u) | P_{0}| f^{(1)}(u) \rangle = g_{1}(f)$. 

We will now perturbatively diagonalize the density matrix terms $\rho_{k,m,n}$. As a warm-up consider the restriction of the problem to the index range $m,n=0,1$, where we can diagonalize the matrices. To 
leading order in $1/2^{k}$ we obtain the eigenvalues: $\rho_{k,0} \approx 1$, $\rho_{k,1} \approx g_{1}(f)/(12 \cdot 4^{k})$, which precisely agrees with the predictions obtained from the expansions of purities (the density matrix eigenvalues are given by $\rho_{k,n} = \Lambda^{2}_{k, n}$). 

Based on this example, we conjecture that eigenvectors are such, that restrictions of the matrix to its lower indices preserve the eigenvalues to leading order. 
Consequently, for the first eigenvector we make the ansatz $e_1 = (1, c_{1}/2^{k}, c_{2}/4^{k}, c_{3}/8^{k},\dots)$, where $c_{s}$ are of order $1$, and do not depend on $k$ to leading order in $1/2^{k}$. Substituting this into the eigenvector equation we indeed find that all coefficients $c_{s}$ can be determined perturbatively in terms of $\langle f^{n}(u) | P_{n-1}| f^{n}(u) \rangle$, while the first eigenvalue is $1$, to leading order.
Similarly, for the second eigenvector we take $e_2 = (d_{0}/2^{k}, 1, d_{2}/2^{k}, d_{3}/4^{k}\dots)$, where $d_{s}$ are again order one coefficients. Substituting into the eigenvector equation allows to find the coefficients $d_{s}$, and the corresponding eigenvalue which, to leading order, is indeed $g_{1}(f)/(12 \cdot 4^{k})$. 
The generalization to the $m$-th eigenvector is as follows: 
\begin{align}
e_m & =(q_{0}/2^{mk}, q_{1}/2^{(m-1)k}, \dots, q_{m-1}/2^{k}, 1, q_{m+1}/2^{k}, \dots) \nonumber \\\ &= (\vec{\tilde{q}}, 1, \vec{\tilde{Q}}),
\end{align}
where we have introduced vectors $\vec{\tilde{q}}, \vec{\tilde{Q}}$ with elements $\tilde{q}_{n} = q_{n}/2^{(m-n)k}, \tilde{Q}_{n} = q_{m+1+n}/2^{k(n+1)}$. We now need to solve the eigenvalue equation:
\begin{equation}
    \hat{\rho}_{k} e_{m} = \rho_{k,m} e_{m},
\end{equation}
where $\rho_{k,m}$ is the $m$-th eigenvalue of the $k$-th qubit reduced density matrix. To this end one can write the matrix $\hat{\rho}_{k}$ in the block form:
\begin{equation}
    \hat{\rho}_{k} = 
    \begin{pmatrix}
        M & l & D \\
        l^{\dagger} & p & w^{\dagger} \\
        D^{\dagger} & w & G
    \end{pmatrix},
\end{equation}
where $M$ is $m \times m$ size matrix, $l$ is a column vector of size $m$, $p$ is a number, w is a column vector, and D, G are matrices. In this notation, the eigenvalue equation may be written in the following form:

\begin{align}
    & M \tilde{q} + l + D \tilde{Q} = \rho_{k,m} \tilde{q}, \\
    & l^{\dagger} \cdot \tilde{q} + p + w^{\dagger} \cdot \tilde{Q} = \rho_{k,m}, \\ 
    & D^{\dagger} \tilde{q} + w + G \tilde{Q} = \rho_{k,m} \tilde{Q}.
\end{align}

We begin solving the system from the second equation. Both $l \cdot \tilde{q} $ and $p$ are of the leading order $1/2^{2 m k}$, while $w^{\dagger} \cdot \tilde{Q} = O(1/2^{(2m+2)k})$. Hence, to the leading order, we can safely neglect the term $w^{\dagger} \cdot \tilde{Q}$. Note also that for consistency $\rho_{k,m} = O(1/2^{2km})$. Consider now the $n$-th equation (with indexation starting from $0$) from the matrix  equation $M \tilde{q} + l + D \tilde{Q} = \rho_{k,m} \tilde{q}$. To the leading order we have $l_{n} = O(1/2^{k(m+n)})$, $M_{n} \tilde{q} = O(1/2^{k(m+n)})$,  $D_{n} \tilde{Q} = O(1/2^{k(m+n+2)})$, $\rho_{k,m} \tilde{q}_{n} = O(1/2^{3km - nk})$ (note that $n < m$). Only first two terms remain, and to the leading order we obtain:
\begin{equation}
    M \tilde{q} = - l, 
\end{equation}
which allows to find the vector $\tilde{q}$. Note that all the factors $1/2^{k}$ cancel out in these equations and we obtain a simple linear system on coefficients $q_{n}$ with $n<m$. The eigenvalue $\rho_{k,m}$ can be found by substituting this solution for $\tilde{q}$ into $\rho_{k,m} = l^{\dagger} \cdot \tilde{q} + p$. Lastly, we should find the vector $\tilde{Q}$. Let us estimate the orders of elements of the $n$-th equation from $D^{\dagger} \tilde{q} + w + G \tilde{Q} = \rho_{k,m} \tilde{Q}$. We have $D^{\dagger}_{n} \tilde{q} = O(1/2^{k(2m+1+n)}$, $w_{n} = O(1/2^{k(2m+1+n})$, $\rho_{k,m} \tilde{Q}_{n} = O(1/2^{k(2m+1+n)})$, while $G_{n} \tilde{Q} = O(1/2^{k(2m+n+3)})$. Hence, we can neglect the term $G \tilde{Q}$ and just obtain $\rho_{k,m} \tilde{Q} = D^{\dagger} q + w$, where all $1/2^{k}$ factors also cancel out to the leading order. 

We can now return to the equation for eigenvalues $\rho_{k,m}$ and further simplify them:
\begin{equation}
    \rho_{k,m} = p - l^{\dagger} M^{-1} l = \frac{\det N} {\det M},
\end{equation}
where the $(m+1) \times (m+1)$ matrix $N$ is defined as follows:

\begin{equation}
    N = 
    \begin{pmatrix}
        M & l \\
        l^{\dagger} & p
    \end{pmatrix}.
\end{equation}

The matrix $N$ is the analog of the matrix $M$ for the eigenvalue $\rho_{k, m+1}$. To simplify the determinants we notice that the matrices $N, M$ can be factorized as follows: $N = F_{m+1} H_{m+1} F_{m+1}, M = F_{m} H_{m} F_{m}$, where $F_{m}$ is a $m \times m$ diagonal matrix with diagonal elements given by:
\begin{equation}
F_{m,nn} = \frac{\sqrt{\langle f^{(n)}(u)| P_{n-1}| f^{(n)}(u) \rangle}}{n! 2^{kn}}
\end{equation}
for $n = 0,1, \dots m-1$, while $H_{m}$ is a $m \times m$ matrix with matrix elements $H_{m,nl} = \frac{1}{n+l+1}$. With these factorizations we obtain the following formula for $\rho_{k,m}$:

\begin{align}\label{eq:eigenvalues}
    \rho_{k,m} &= \left(\frac{\det{F_{m+1}}}{\det{F_{m}}}\right)^{2} \times \frac{\det{H_{m+1}}}{\det{H_{m}}} \nonumber \\ &= \frac{\langle f^{(m)}(u)| P_{m-1}| f^{(m)}(u) \rangle}{m!^{2} 4^{km}} \times \frac{\det{H_{m+1}}}{\det{H_{m}}}.
\end{align}
Here the first factor contains information on the function and the $1/4^{km}$ scaling, while the factor $\frac{\det{H_{m+1}}}{\det{H_{m}}}$ is just a positive numerical coefficient, independent of $k$ and of the details of the function. 

\qed

\subsection{Proof of Corollaries \ref{cor:eigenvalue} and \ref{cor:entropy}}

We prove both results simultaneously.

Let the subleading part of the entanglement spectrum be of order $s$. Hereafter, we can make additional assumption that the third in size entanglement coefficient is at least of order $O(s^{2})$. This assumption follows from Theorem \ref{thm:density_matrix}. Then from the state normalization $\sum_{i} \Lambda^{2}_{k,i} = 1$, it follows that $\Lambda_{k,0} = 1 - s^{2}/2 + O(s^4)$. Using Eq.\ref{eq:purspect} the purity has now the form:
\begin{equation}
    p_{k} = (1-s^{2}/2 + O(s^{4}))^{4} +O(s^{4}) = 1 - 2 s^{2} + O(s^{4}).
\end{equation}
Comparing with the expansion of purity in $1/2^{k}$ in Eqs.\ref{eq:pk1}, \ref{eq:pk2} we obtain the result of the first corollary.

To prove the result for the entanglement entropy it's enough to compute:
\begin{align}
S_{k} &= -(1-s^{2})\log{(1 - s^{2})} - s^{2} \log{s^{2}} \nonumber \\ & \approx s^{2}(1-\log{s^{2}}) \sim Ck/4^{k}
\end{align}
for some constant $C$ which one can compute explicitly from other terms.

Note that in both results we used an assumption that $s$ is small enough and asymptotic expansions can be applied. The latter is supported by Theorem~\ref{thm:purities} for large enough $k$.
\qed

\subsection{Breaking the smoothness assumption}\label{subsec_smoothness}
Here we briefly comment on how our results change when the smoothness assumption is not satisfied. Consider the case where the function $f$ has only $l$ absolutely continuous derivatives. Then, by Taylor's theorem it can be expanded in series: $f(u+v/2^{k}) = f(u) + f'(u)v/2^{k} + ... + f^{(l)}(u) v^{l} /l!2^{k l} + R_{l} (u, v)$, where $R_{l}$ is the remainder term in the integral representation: $R_{l}(u,v) = \int_{0}^{v} f^{(l+1)}(u+z/2^{k}) (v-z)^l dz / l! 2^{k(l+1)}$. This remainder term is is of the order of $O(1/2^{k(l+1)})$, and hence it cannot spoil the hierarchy of matrix elements in the density matrix. As a result, all the results which depended only on hierarchy of magnitudes of the matrix elements, like the formula for the leading entanglement spectra, will not change. In particular, the first $(l+1)$ density matrix eigenvalues will follow Eq.~\ref{eq:eigenvalues}. In numerical experiments we observed an even stronger result, where the first $l+2$ eigenvalues behaved followed the scaling derived for the smooth functions. This may be connected to the fact that piecewise polynomials do have exact MPS representations of low rank. We leave a more detailed study of entanglement of non-smooth functions to future research.

\section{Examples, expansion constants and probability distributions}
\label{app:B}

\subsection{Analytical examples for the theoretical smoothness results}

The expansions in $1/2^{k}$ from Theorem~\ref{thm:purities} are valid when $g_{1}(f)/(6\times4^{k}) < 1$ (\emph{i.e.}~when the second expansion term in Eq.~\ref{eq: purity} is smaller then $1$). To shed light on the intuitive understanding of this condition consider the example of a Gaussian $f(x) = \exp{[-(x-0.5)^{2} /4 \sigma^{2}]}$. Then $g_{1}(f) = \langle (x-0.5)^{2} /4 \sigma^{4} \rangle = 1/4 \sigma^{2}$ (we ignored the interval finiteness, which is valid for low enough $\sigma$) and the validity condition reduces to $k > \log_{2}{1/2\sigma} - \log_2{6}$. As $2 \sigma$ is the characteristic length scale of the Gaussian the condition implies that the asymptotic results are valid for qubits $k$ responsible for scales $1/2^{k}$ smaller than the characteristic length scale of the distribution. 
Thus, while for small $k$, corresponding to large scale function variations, entanglement entropies and other measures can be large, for large $k$ (small scale variations) they decay exponentially in $k$. 

Next consider the function $f(x) = sin(\pi x)$, for which the integrals in the density matrix definition can be computed exactly in full \cite{Oseledets_2012}. The density matrix is given by:
\begin{multline}
    \rho_{k} = |\sin{\pi x} \rangle \left(\frac{1}{2}+ \frac{2^{k}}{4\pi} \sin(2 \pi /2^{k})\right) \langle \sin{\pi x}| +\\ + |\cos{\pi x} \rangle \left(\frac{1}{2}-\frac{2^{k}}{4\pi} \sin\left(2 \pi / 2^{k}\right)\right) \langle \cos{\pi x}| + \\ + \frac{2^{k}}{2\pi} \sin^{2}{(\pi/2^{k})} (|\cos{\pi x} \rangle \langle \sin{\pi x}| + |\sin{\pi x} \rangle \langle \cos{\pi x}|).
\end{multline}
Then
\begin{multline}
   \rho_{k} = |\sin{\pi x} \rangle (1+2^{k} \sin(2 \pi / 2^{k})/(2 \pi))/2 \langle \sin{\pi x}| + \\ + |\cos{\pi x} \rangle (1-2^{k} \sin(2 \pi / 2^{k})/(2 \pi))/2 \langle \cos{\pi x}| + \\ + 2^{k} \sin^{2}{(\pi/2^{k})}/(2 \pi) (|\cos{\pi x} \rangle \langle \sin{\pi x}| + |\sin{\pi x} \rangle \langle \cos{\pi x}|).
\end{multline}
Since $|\sin{\pi x} \rangle$ and $ |\cos{\pi x} \rangle$ are orthogonal for all $k$ and have the same norm we do not need the Gram-Schmidt procedure. 
Diagonalizing the density matrix we obtain the following eigenvalues: 
\begin{equation}
\rho_{k,\pm} = \frac{1}{2} \pm \frac{2^{k}}{2 \pi} \sin{(\pi/2^{k})}.
\end{equation}
Expanding the smaller one we obtain to leading order $\pi^{2}/(12 \cdot 4^{k})$, which precisely agrees with the Theorem~\ref{thm:density_matrix}, with $g_{1}(f) = \pi^{2}$. Note, that entanglement spectra decrease with $k$, so for large enough $k$ we can truncate the bond dimensions to $1$. 

\subsection{The third eigenvalue and their sum}

For the error analysis we compute the expression for the third entanglement coefficient $\Lambda_{k,2}$. To the leading order for large enough $k$ we have:
\begin{equation}
    \Lambda_{k,2} \approx \sqrt{\frac{\langle f^{(2)}(u) | P_{1}| f^{(2)}(u) \rangle}{720 \times 16^{k}}} = \sqrt{\frac{g_{2}(f)}{720 \times 16^{k}}},
\end{equation}
where we defined $g_{2}(f) = \langle f^{(2)}(u) | P_{1}| f^{(2)}(u) \rangle$. 

The sum of the remaining eigenvalues appear in computing fidelities between an ideal state and a circuit's state after a single layer of the analytical decomposition.
For the entanglement spectra, from Theorem~\ref{thm:density_matrix} we know that $\Lambda_{k,i} = O(1/2^{k i})$. Using these scaling relations we can derive the leading estimate of the fidelity after one layer of a generalized analytic decomposition:
\begin{equation}
    \sum_{\substack{k\geq m\\i\geq2}} \Lambda_{k,i}^{2} \approx g_{2}(f) /720 \times 15\times 16^{m-1}  + \dots,
\end{equation}
where $m$ is the first bond with truncation. It can be the second bond (the first bond always has the bond dimension 2 and is not truncated) or some larger bond if we use analytic decomposition with three-site gates for the first qubits. Note also, that for one-layer analytic decomposition, the resulting fidelity does not depend on the position of the canonical form center, as it does not change the error of truncation. 

\subsection{Exact functional coefficients for probability distributions}
Here we give explicit expressions for the $g_{1}(f)$ functions for the symmetrical normal ($\mu=L/2$), log-normal and L\'evy distributions, obtained by performing the integrals in Eq.\ref{eq:pk2} for these examples:

\begin{align}\label{eq:g1normal}
    g_{1}&(\sqrt{p(x)_{normal}}) \\ \nonumber &= \frac{L^{2}}{4 \sigma^{2}} \left[1 - \sqrt{\frac{L^{2}}{2 \sigma^{2}}} \frac{\exp{\left(-\frac{L^{2}}{8 \sigma^{2}} \right)}} {\left( \Gamma(1/2) - \Gamma(1/2, L^{2}/8\sigma^{2})\right)}\right]
\end{align}

\begin{multline}\label{eq:g1lognormal}
    g_{1}(\sqrt{p(x)_{log-normal}}) = \\= \frac{L^{2} \exp{\left(-2\mu + 2 \sigma^{2}\right)}}{4 \sigma^{2} \left[ \sqrt{2 \pi} - \frac{1}{\sqrt{2}}\Gamma \Bigl (\frac{1}{2}, \frac{((\log L -\mu)/\sigma)^{2}}{2} \Bigr )  \right]} \times \\ \times \biggl [\frac{1 +\sigma^{2}}{\sqrt{2}} \biggl (2\sqrt{\pi} - \Gamma \biggl (\frac{1}{2}, \frac{((\log L -\mu)/\sigma + 2 \sigma)^{2}}{2} \biggr ) \biggr ) + \\ + \left((2-\sqrt{2}) \sigma -\frac{\log L -\mu}{\sqrt{2}\sigma} \right) \exp{[\frac{(\frac{\log L -\mu}{\sigma} + 2 \sigma)^{2}}{-2}]} \biggr ] - \\ - \frac{L^{2}\exp{\{-2\mu + \sigma^{2} - ((\log L -\mu))/\sigma + \sigma)^{2}\}}}{2\sqrt{2} \sigma^{2} \left[2\sqrt{\pi} - \Gamma \Bigl (\frac{1}{2}, \frac{((\log L -\mu)/\sigma)^{2}}{2} \Bigr) \right]^{2}}
\end{multline}

\begin{multline}\label{eq:g1levy}
    g_{1}(\sqrt{p(x)_{Levy}}) = \\ = \frac{21 L^{2}}{8 c^{2}} + \frac{4 L^{2} \exp{[-c/2L}]}{c^{2} \Gamma(1/2, c/2 L)} \left[ \frac{21}{16} \left(\frac{c}{2 L} \right)^{1/2} + \frac{7}{8} \left(\frac{c}{2 L} \right)^{3/2}+ \right. \\ +\left. \frac{1}{8} \left(\frac{c}{2 L} \right)^{5/2} + \frac{1}{4} \left(\frac{c}{2 L} \right)^{7/2}\right] + \\ - \frac{L^{2}}{c^{2} \Gamma(1/2, c/2L)^{2}} \left(\frac{c}{2 L}\right)^{3} \exp{[-c/L]}
\end{multline}

For the L\'evy distribution we computed $g_{1}(f)$ for the discretization on the finite interval $L$. The leading term $21 L^{2} / 8 c^{2}$ defines the characteristic scale $k = \log{L/c}$, where the entanglement spectra become small. Other terms define the subleading corrections due to the finiteness of the interval. Here $\Gamma(s, x)$ is the incomplete $\Gamma$ function, which comes from the normalization of the L\'evy distribution on the finite interval.

Analogous explicit formulae can also be derived for the $g_{2}(f)$ coefficient. Here we give the one for the same normal distribution (the others result in impractically large expressions):

\begin{multline}\label{g2normal}
    g_{2}(\sqrt{p(x)_{normal}}) = \\ = \frac{L^{4}}{8 \sigma^{4}} - \frac{L^{4}}{4 \sigma^{4}} \left(\sqrt{\frac{L^2}{32 \sigma^{2}}} + \sqrt{\frac{L^{2}}{8 \sigma^2}}^{3}\right)\frac{\exp{\left[-\frac{L^2}{8 \sigma^2}\right]}}{\Gamma(
    \frac{1}{2}
    ) - \Gamma(\frac{1}{2}, \frac{L^2}{8 \sigma^{2}})} - \\ - \frac{L^{6}}{32 \sigma^6} \frac{\exp{\left[-\frac{L^2}{4 \sigma^2}\right]}}{[\Gamma(\frac{1}{2}) - \Gamma(\frac{1}{2}, \frac{L^2}{8 \sigma^{2}})]^{2}}
\end{multline}
If we drop the finitness of the interval, that is, consider the large $L$ limit, the expression simplifies to:
\begin{equation}
    g_{2}(\sqrt{p(x)_{normal}}) = \frac{1}{8\sigma^{4}}.
\end{equation}

\section{Entanglement and function localization}
\label{app:C}

In Sec.\ref{sec:generalres} we showed that the entanglement on the smallest scales exhibits universal exponential decay. We can similarly ask how the entanglement behaves on the qubits corresponding to the largest scales. 

It is easy to see that this behaviour \emph{cannot be universal}, by the following construction: for any given quantum state on $n<N$ qubits one can convert its amplitudes into the corresponding tensor of $2^n$ discrete ``function values", which can further be polynomially interpolated to yield a smooth function. This function can have an arbitrarily complicated behaviour down to the scale $L/2^n$. On the other hand a function can equally well be arbitrarily simple on large scales, such as for instance in the case of a periodic function with a period $T$. It is easy to see that the qubits responsible for the larger scales $L'$, \emph{i.e.}~$T<L'<L$, will be in an equal weight superposition (a product of $|+\rangle$ states), and completely disentangled from the qubits corresponding to the smaller scales. These two opposite extremes show that there is no universality for large scales.

While we do not provide here rigorous results, it is possible to characterize the large scale entanglement slightly better for specific classes of functions.
An important case are the \emph{localized} functions. Informally, a function is localized exponentially or polynomially if the integral of the function over increasing subsets containing the domain of localization approaches unity (we assumed normalized function) with the remainder decaying polynomially or exponentially in the size of the subsets.

We can follow an analogous path to talk about localization on a finite domain, but the presence of a binary grid, which we work with, introduces additional subtleties. For the purpose of clarity, and at the expense of full generality, we restrict the discussion to a localization on a sub-interval $I_m$ of size $L/2^{m}$ (and scale $m$), which is ``compatible" with the binary grid. We define a compatible interval $I_{\sigma_1 \ldots \sigma_m} \equiv I_m$ to be the set of all points sharing a fixed first $m$ bits $\sigma_1 \ldots \sigma_m$ of the binary expansion. A sequence of increasing intervals $I_m \subset I_{m-1} \subset \ldots \subset I_1$ exists, defined by freeing the successive digits in the binary expansion. For the function localized exponentially (resp.~polynomially) to $I_m$ we now have:
\begin{equation}
\int_{I_{l}}|f(u)|^{2} du \approx 1 - C(|I_{l}|),
\end{equation}
for $l=m,\ldots,1$,
where the correction $C(|I_{l}|)$ decreases exponentially (polynomially) in the size of $I_{l}$.

Note now that $\int_{I_{l}}|f(u)|^{2} du$ is in fact (see Eq.\ref{eq:eq33}) the diagonal element of the reduced density matrix $\rho_{l, u_{l} u_{l}}$, with $u_l \equiv \sigma_1 \ldots \sigma_l$. Thus $\rho_{l, u_{l} u_{l}}$ is exponentially or polynomially close to one, for localized functions. Conversely, the off-diagonal elements of the reduced density matrix by Eq.\ref{eq:eq33} will be exponentially/polynomially suppressed, which immediately implies that the entanglement across bonds $m,\ldots,1$ is also similarly suppressed (see Fig.\ref{fig2: localization}). A delta distribution is an extreme example of this: all but on elements of the density matrix vanish, and the distribution can be represented as a product state.
Thus, localization properties influence the behaviour of the entanglement on the bonds representing the largest spatial scales.

\begin{figure}[h]
\centering
\includegraphics[width= \columnwidth]{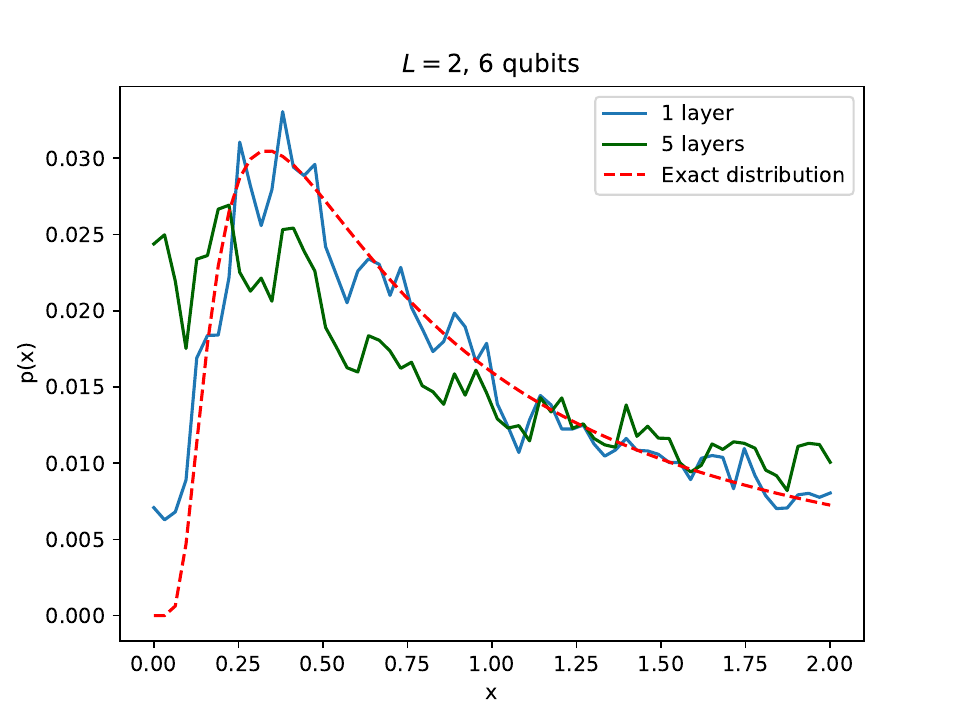}      
\caption{\label{fig:Levy_scaled_noise} The L\'evy distribution encoded with 1- and 5-layer circuits on a noisy simulated quantum device (IBM \texttt{FakeToronto}). The output of the theoretically more accurate but deeper circuit is much stronger affected by the noise. Such trade-off effects are very important in practice.
       }
\end{figure}

\section{Accuracy-depth trade-off in the presence of noise}\label{app:noisy_execution}
As discussed in the main text, noise induces a practical trade-off, where theoretically better but deeper circuits result in worse approximation of the distribution. To probe this we performed noisy simulations on the IBM \texttt{FakeToronto} backend. We compare the results of the execution of a 1-layer and 5-layer encoding circuits in Fig.~\ref{fig:Levy_scaled_noise}. While in the absence of noise the latter is more accurate, with noise the distribution features are largely destroyed, while the nominally less accurate 1-layer encoding performs much better.

\end{document}